\documentclass[%
 aps,twocolumn,
 amsmath,amssymb,superscriptaddress, 
prb
]{revtex4-1}

\usepackage{graphicx}
\usepackage{dcolumn}
\usepackage{bm}
\usepackage{color,soul}
\usepackage[usenames,dvipsnames]{xcolor}
\usepackage[colorlinks=true,linkcolor=RoyalBlue,citecolor=OliveGreen,urlcolor=OliveGreen,linktoc=page]{hyperref} 

\usepackage{enumerate}
\usepackage{grffile}
\usepackage{balance}
\usepackage{float}

\DeclareMathOperator{\sech}{\text{sech}}
\newcommand{\avg}[1]{\langle #1 \rangle}
\newcommand{\bavg}[1]{\left \langle #1 \right \rangle}

\begin{document}

\title{Phase Slips in Superconducting Weak Links}

\author{Gregory  Kimmel}
\affiliation{Department of Engineering Sciences and Applied Mathematics, Northwestern University, 2145 Sheridan Road, Evanston, Illinois 60202, USA
}
\affiliation{Materials Science Division, Argonne National Laboratory, 9700 South Cass Avenue, Argonne, Illinois 60439, USA}
\author{Andreas Glatz} 
\affiliation{Department of Physics, Northern Illinois University,  DeKalb, Illinois 60115, USA }
\affiliation{Materials Science Division, Argonne National Laboratory, 9700 South Cass Avenue, Argonne, Illinois 60439, USA}
\author{Igor S. Aranson}
\affiliation{Materials Science Division, Argonne National Laboratory, 9700 South Cass Avenue, Argonne, Illinois 60439, USA}
\affiliation{Department of Engineering Sciences and Applied Mathematics, Northwestern University, 2145 Sheridan Road, Evanston, Illinois 60202, USA
}

\date{\today}
\begin{abstract} 
Superconducting vortices and phase slips are primary mechanisms of dissipation in superconducting, superfluid, and cold atom systems. While the dynamics of vortices is fairly well described, phase slips occurring in quasi-one dimensional superconducting wires still elude understanding. The main reason is that phase slips are strongly non-linear time-dependent phenomena that cannot be cast in terms of small perturbations of the superconducting state.  Here we study phase slips occurring in superconducting weak links. Thanks to partial suppression of superconductivity in weak links, we employ a weakly  nonlinear approximation for dynamic phase slips. This approximation is not valid for homogeneous superconducting wires and slabs. 
Using the numerical solution of the time-dependent Ginzburg-Landau equation and bifurcation analysis of stationary solutions, we show that the onset of phase slips occurs via an infinite period bifurcation, which is manifested in a specific voltage-current dependence. Our analytical results are in good agreement with simulations. 
\end{abstract} 

\maketitle

\section{\label{intro}Introduction} \label{sec:level1}
The motion of Abrikosov vortices is recognized as the main cause of dissipation in type-II superconductors \cite{blatter}. Conversely, in thin nanowires, the motion of vortices is impeded and phase-slip events are responsible for the dissipation. Phase slips, changing the phase difference of the superconducting order parameter by $2 \pi$, may be caused by different physical mechanisms. Thermally activated phase slips at high temperatures and small applied currents are well understood \cite{tinkham1996introduction}. At very low temperatures, phase slips can be caused by quantum fluctuations (aptly called quantum phase slips) \cite{mooij2006superconducting,lau2001quantum,glatz+prl02}. Phase slips are not unique to superconductors, they also occur in superfluid systems \cite{anderson1966considerations,langer1967intrinsic, schwarz1990phase}, and more recently, dissipation due to phase slips were studied in cold atom systems \cite{mckay2008phase,scherpelz+prl14,scherpelz+pra15}. In particular, phase slips can be triggered in a superfluid cold atom system by a rotating weak link \cite{wright2013driving}. 

Even without thermal and quantum fluctuations, the phase slip phenomena and dissipative (or resistive) states can be induced by an applied current \cite{Tinkham1974PSC,Meyer1974Voltage}. Magnetic field penetrates type-II superconductors in the form of Abrikosov vortices. If an external current is applied, the Lorentz force induces motion of the vortices. This motion is the main cause of dissipation in 2D and 3D superconductors. However, in quasi-one dimensional nanowires with the coherence length $\xi(T)$ and the penetration depth $\lambda(T)$ large compared to the wire diameter, vortex motion is suppressed. In this situation the transition to the normal state was made through successive voltage jumps which are attributed to the appearance of phase slip centers\cite{Tinkham1974PSC,Meyer1974Voltage}. A study of this phenomenon was given first by Kramer and Baratoff who found that slightly below the depairing current, there is a dissipative state which consists of localized phase slips occurring in the superconducting filament \cite{Quasi-1D-Superconductors}. In a narrow range of currents close to the depairing current, the material is superconducting except in narrow regions where phase slip centers (PSCs) occur. The period of these PSCs diverge as the external current approaches the lower bound in this narrow region. It was also shown that random thermal fluctuations allow for phase slips \cite{Little-Theory}, but these did not persist indefinitely. Further numerical study of the one-dimensional time-dependent Ginzburg-Landau equation revealed periodic phase slips existing in a narrow range of currents close to the depairing current \cite{kramer1984structure,rangel1989theory}. Follow-up numerical studies of narrow two-dimensional superconducting strips discovered a transition from a phase-slip-line to vortex pairs \cite{weber1991dissipative}. Periodic lattices of the phase slip centers were studied in the context of vortex penetration in thin superconducting films near the third critical magnetic field \cite{aranson1998surface}.
Using a saddle-point approximation for the Ginzburg-Landau energy in narrow superconducting strips,  the dependence of  voltage drop vs temperature and bias current  (neglecting thermal fluctuations) 
was studied in [\onlinecite{ovchinnikov2015phase}].

The situation is different, however, for spatially inhomogeneous systems, such as superconductors with macroscopic defects or weak links \cite{LA-Theory}. Perhaps the most famous examples are Dayem bridges and Josephson junctions \cite{Josephson,Dayem}. The mechanism for dissipation in these cases is the quantum tunneling of Cooper pairs between the two superconductors, which is caused by a phase difference between the weakly-linked superconductors. When the current is below some threshold $j_c$, the phase difference is fixed in time and a stationary superconducting state persists. Above this threshold, the solution exhibits oscillations, which lead to a finite voltage. In a review paper by Ivlev and Kopnin, inhomogeneities were analyzed, but in regards to the stability of the normal state \cite{Kopnin}. Thus, their analysis involved currents much closer to the GL critical threshold $j_{GL} = 2/\sqrt{27}$. A lower bound $j_1$ at which the normal state was globally unstable (i.e. arbitrary small perturbation lead to instability of the normal state), and above which there was a critical-sized perturbation which separated the normal and superconducting states was estimated. Also, an upper critical current $j_2$ such that the normal state was absolutely stable for an external current $j_0 > j_2$ was found. An inhomogeneity much smaller than the coherence length, $\xi(T)$, was used and was approximated by a $\delta$ function, simplifying the algebra. Here we consider a more realistic situation for the type-{II} high-temperature superconductors:  an inclusion on the scale of $\xi(T)$. The transition we are interested in analyzing, occurs between the non-uniform superconducting state and the oscillatory state with phase slips. Therefore, the steady state and linearization in this paper are much more complex then in analyzing the normal state. The authors of [\onlinecite{van1981superconductor}] have shown experimental results of weak-links with non-hysteric behavior.

The phase slip state of homogenous systems have recently been analyzed in much greater detail \cite{PhysRevB.84.094527}. Using bifurcation analysis, Baranov et. al. extract the normal form of a saddle-node bifurcation when the current is near the critical current. They then correctly determine the characteristic scaling law and show its agreement with numerical simulations. The period diverges in an infinite-period saddle-node bifurcation as $j_0 \to j_c$. These authors further expanded upon their analysis by showing the important role that the material parameter $u$ plays in the type of bifurcation that can occur \cite{PhysRevB.87.174516} ($u$ is related to the electric field penetration depth). They observed that for finite lengths and values of $u$ above some critical threshold $u_{c2}$, numerical simulations showed hysteresis in the I-V curve. However, our work focuses on analytical methods for the inhomogeneous system, which as stated previously makes the steady state and linearization to much more difficult to handle. We show that a simplified system can be obtained through weakly nonlinear analysis and that this system contains the normal form obtained in [\onlinecite{PhysRevB.84.094527}] as the size of the weak link shrinks to zero. We also demonstrate that in addition to the infinite period bifurcation for small $u$, a  hysteresis exists in our system for large $u$ values, similar to that in Ref. [\onlinecite{PhysRevB.84.094527}]. However, in contrast to previous studies, our reduced two-dimensional nonlinear system exhibits evolution of periodic orbits and a transition between superconducting and normal states that are not properly captured by the one-dimensional model in Ref [\onlinecite{PhysRevB.84.094527}]. 

A work by Michotte et. al. in [\onlinecite{PhysRevB.69.094512}] have found that the condition for PSCs to occur is based on the competition between two relaxation times: the relaxation time for the magnitude of the order parameter $t_{|\Psi|}$ and the relaxation time for the phase of the order parameter $t_\phi$. They observed that phase slips are possible only when $t_\phi < t_{|\Psi|}$. A linearized Eilenberger equation in the dirty limit was studied, resembling a generalized TDGL equation with additional parameters related to inelastic electron-phonon collisions, which was first derived in [\onlinecite{PhysRevLett.40.1041}]. They derived an approximate critical current via this equation and their results implied that there was a finite maximal oscillation period for the order parameter. In contrast, for weak links all oscillation periods diverged. The generalized GL equation used contained an additional parameter $\gamma$ characterizing relative superconducting phase relaxation time  (for us, $\gamma = 0$). For large $\gamma$ values  hysteresis was observed in the I-V curve. On a qualitative level, the effect of increasing parameter $\gamma$ is similar to an increase in parameter $u$ \cite{PhysRevB.84.094527}. Correspondingly, we observed hysteresis when $u \gg 1$. The authors of [\onlinecite{berdiyorov2012dynamic}] have done numerical analysis of a periodic array of weak links using the generalized TDGL equation. They showed I-V curves for different magnetic fields, however no analysis of the divergence of the period of vortices was presented.

We focus on a 1D superconductor, separated by a normal or weakly superconducting inhomogeneity. The complete system is modeled by a spatially dependent critical temperature $T_c(x)$.  The weak link is created by a lower transition temperature inside an interval $I=[-r,r]$, which leads to a suppression of the order parameter. Here $r$ is the inclusion radius. Below some critical current, this system relaxes to a stationary superconducting state, but above it, the superconductor exhibits a finite voltage with oscillatory behavior. Thermal fluctuations are initially not considered in this model and therefore does not cause a finite voltage in the superconducting state. The Josephson junction analysis is not applicable here. Indeed, since there is no dielectric contact between the two superconductor pieces, the phase should always be the same, implying zero voltage. We will show via simulations of the time-dependent Ginzburg-Landau equation, that the oscillations in the voltage is caused by phase slips in the center of the inclusion. The system approaches this state via a saddle-node bifurcation of two superconducting states, which occur at the critical current 
(at a saddle-node bifurcation stable and unstable stationary superconducting states annihilate and a periodic resistive state appears). The suppression of the order parameter in and near the weak link allows us to employ analytical methods in the vicinity of the critical current. We derive a reduced two-dimensional system governing the time evolution of the phase slip solution and describe a sequence of transitions between superconducting and dissipative states. 

The paper is organized as follows: section \ref{sec:level2} describes the model, section \ref{sec:level3} deals with the stationary case and estimates the critical current which is obtained from the saddle-node bifurcation condition. Sections \ref{sec:level4}-\ref{sec:level7} deal with the time periodic solutions, extracting a time-dependent system via weakly nonlinear analysis and then studying the simplified model to show that it exhibits the same qualitative behavior. In section \ref{sec:discussion}, we interpret our analytical results, show the correspondence to the parameters of the superconductor and its effects on the phase slip state. Finally, section \ref{sec:conclusion} gives closing remarks and ideas for further study.

\section{\label{sec:level2}Governing equations}
The time-dependent Ginzburg-Landau equations (TDGLE) are obtained by minimization of the GL free energy\cite{aranson2002world}. 
In the absence of a magnetic field this results in
\begin{align} \Gamma \left(\partial_t + i \frac{2 e}{\hbar} \mu \right) \Psi = a_0 \nu(x) \Psi - b |\Psi|^2 + \Psi + \frac{\hbar^2}{4m} \partial_x^2\Psi, \label{Igor} \end{align}
where $\Gamma,a_0,b$ are phenomenological parameters that can be found from the microscopic theory~\cite{BCS-Theory}, $e,m$ are the electron charge and mass, $\mu$ is the scalar potential, and $\nu(x)$ a spatially dependent linear coefficient modeling inhomogeneities in the system. Following Sadovskyy et al.\cite{sadovskyy2015stable}, we define the $+x$ direction be the direction of the external current and obtain the following dimensionless form:
\begin{subequations}
\begin{equation}
u(\partial_t + i \mu) \Psi =  \partial_x^2\Psi + [\nu(x) - |\Psi|^2]\Psi \label{GL1} 
\end{equation}
with the total current $j_0$
\begin{equation}
j_0 = \Im(\Psi^*  \partial_x\Psi) -  \partial_x\mu. \label{GL2}
\end{equation}
\end{subequations}
Here $\Psi$ is the complex order parameter, satisfying $|\Psi| = 1$ in the purely superconducting state, and $|\Psi| = 0$ in the normal state. The parameter $u = \Gamma/a_0 \tau_{\text{GL}}$ with time $\tau_{\text{GL}} = 4 \pi \sigma \lambda_0^2/c^2$, $\lambda_0 = \sqrt{\frac{m c^2}{8\pi e^2 \psi_0^2}}$ is the magnetic penetration depth ($c$ the speed of light) and $\psi_0 = \sqrt{a_0/b}$ is the equilibrium value of the order parameter when spatial variations are neglected, i.e., $\nu(x)=1$. The zero temperature coherence length $\xi_0 = \sqrt{\frac{\hbar^2}{4m a_0}}$ is used for the unit of length. For more details see [\onlinecite{sadovskyy2015stable}].

We apply periodic boundary conditions for $\Psi$. Since $\mu$ is on average an increasing function of $x$, there is necessarily a discontinuity at the boundary. This is resolved by making the following transformations:
\begin{subequations}
\begin{align}
\Psi & = \tilde \Psi e^{i K(t) x} \\
\mu & = -Ax + \tilde \mu.
\end{align}
\end{subequations}
Here, $\tilde \mu$ is a periodic function in $x$. Essentially, we are moving the growth of $\mu$ to the phase of $\Psi$. The growth in $K$ now does not affect the magnitude. Indeed, this also allows us to \emph{rewind} $K$ through $K \to K - (2\pi/\Delta x) \lfloor K \Delta_x/2\pi \rfloor$ which will remove any error from rapid phase  oscillations \cite{sadovskyy2015stable}. Inserting this into \eqref{GL1} gives
\[ u[\partial_t + i x( \partial_t K - A) + i \tilde \mu] \tilde \Psi = (\partial_x + i K)^2 \tilde \Psi + [\nu(x) - |\tilde \Psi|^2] \tilde \Psi. \]
Setting $\partial_t K = A$ eliminates the linear term. Now inserting this into \eqref{GL2}, we have
\[ j_0 = \Im(\tilde \Psi^* \partial_x\tilde \Psi) + |\tilde \Psi|^2 K + \partial_t K - \partial_x\tilde \mu. \]
Averaging this equation over space and noting that $\avg{\tilde \mu_x} = 0$ results in an ordinary differential equation (ODE) for $K$
\[ \partial_t K + \bavg{|\tilde \Psi|^2} K = j_0 - \Im \bavg{\tilde \Psi^* \partial_x \tilde \Psi} \equiv j_n. \]
For clearer notation, we now suppress the tildes, and we arrive at our modified TDGLE
\begin{subequations} \label{NGL}
\begin{align}
u(\partial_t + i \mu) \Psi & = (\partial_x + i K)^2 \Psi + [\nu(x) - |\Psi|^2] \Psi \label{NGL1} \\
\mu_x & =  \Im(\Psi^* \partial_x\Psi) + \partial_t K + |\Psi|^2 K - j_0 \label{NGL2} \\
j_n & = \partial_t K + \bavg{|\Psi|^2} K \label{NGL3}.
\end{align}
\end{subequations}

The integration domain is periodic with the period $L$.  For the numerical integration, we generally took $L = 20$ and $u = 1$, however this was relaxed to see if the qualitative behavior changed. We verified that increasing $L$ does not affect the results, however changing $u$ can have a large effect (see section \ref{subsec:d2}). To make the analysis simpler, we placed the weak link of length $2r$ symmetrically at the origin in the interval $I$. The inclusion's effect enters through the term $\nu(x) \Psi$ defined by
\begin{equation} \label{T(x)}
\nu(x) \equiv \begin{cases} 1, & x \not \in I \\ -C, & x \in I \end{cases}\,.
\end{equation}
Numerical analysis has shown that for $L \gg r$ there exists a critical current $j_c$, which is a function of $r$ that separates the dynamics of this system. For $j_0 < j_c$, the system goes to a stationary superconducting state, while for $j_0 > j_c$ the system exhibits a dissipative state represented by periodic phase slips occurring in the center of the inclusion via a stable limit cycle. In the following sections, we explain these results analytically. We first provide an analytical approximation of the critical current. Next, we extract a coupled two-dimensional nonlinear system of ODEs from \eqref{GL1} which describes qualitatively, the correct behavior for suitable choices of the coefficients of the simplified system.

\section{\label{sec:level3}The stationary case $j_0 < j_c$}
In the superconducting state with an applied current of $j_0 < j_c$, it can be shown that $\mu = 0$, (see appendix \ref{nomu} for details). To proceed, we rewrite  \eqref{GL1} in terms of amplitude and phase of the order parameter, i.e., $\Psi = F e^{i \phi}$. Inserting this into \eqref{GL1} and \eqref{GL2} gives for the stationary equation 
\begin{subequations} \label{eq:stationary and current}
\begin{align}
0 & =  \partial_x^2F + [ \nu(x) -  (\partial_x\phi)^2 - F^2] F \label{stationary} \\
j_0 & = F^2  \partial_x\phi. \label{current}
\end{align}
\end{subequations}
Plugging \eqref{current} into \eqref{stationary} gives the nonlinear ODE
\begin{equation} \label{stat2}
0 =  \partial_x^2F + [\nu(x) - j_0^2 F^{-4} - F^2]F.
\end{equation}

\subsection{Large $C$ approximation} 

\begin{figure}[htb]
\vspace*{-0.2in}
\includegraphics[width=\columnwidth]{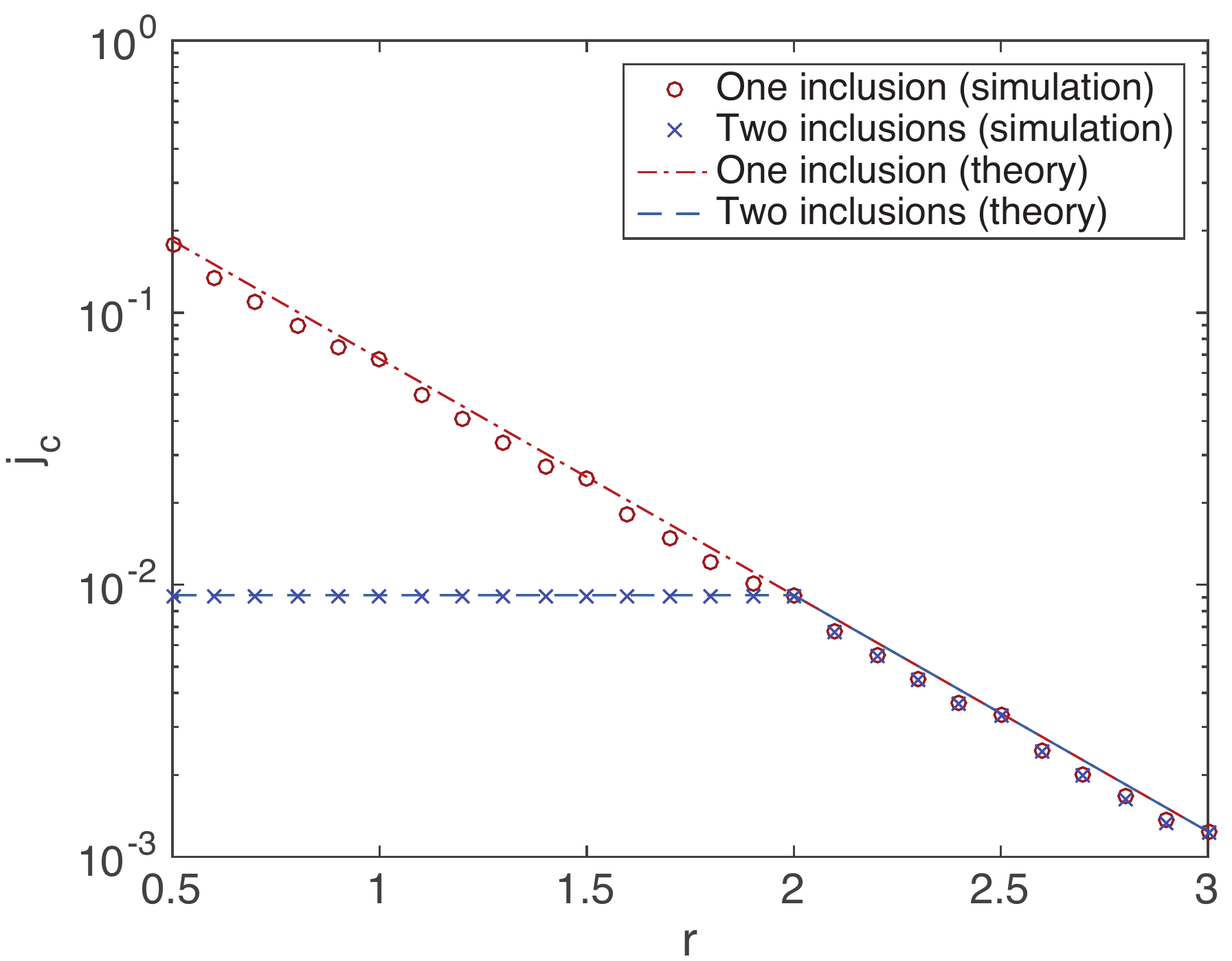}
\vspace*{-0.2in}
\caption{The critical current as a function of inclusion size using \eqref{eq:critcurrent} (e.g. $C = 1$ with \eqref{critcurrentapprox}). For the two inclusions, one inclusion is held fixed at $r = 2$. Above the curves the superconducting order parameter $\Psi$ oscillates.}
\label{fig:critcurrent}
\end{figure}

We now assume a large $C$ approximation, that is, the weak link strongly suppresses superconductivity in the inclusion (i.e. $C \gg j_0^2 F^{-4}$). This allows us to neglect the nonlinear term and obtain a first order approximation of the solution of \eqref{eq:stationary and current}. From this we notice that \eqref{stationary} has a first integral for both the inclusion domain and the superconducting domain. Asymptotic analysis of the size of these coefficients gives us a condition for $j_c$ given by
\begin{equation} \label{critcurrentapprox}
j_c = \frac{1}{2 \sqrt C} e^{-2r \sqrt C},
\end{equation}
for details see appendix \ref{critcurrent}. Setting $C = 1$, we have that
\begin{equation} \label{eq:critcurrent}
j_c = \frac{1}{2} e^{-2r}.
\end{equation}
Comparing this approximation with numerical simulations, we see that the large $C$ approximation with $C = 1$ is in good agreement with the numerical solution (see Fig. \ref{fig:critcurrent}). Thus, we derived that a weak link results in a exponential suppression of the critical current as a function of the inclusion width $2r$ and strength $C$. A similar result was obtained through a different method in Ref. \onlinecite{rangel1989theory}. However, our method is appealing for the simple generalization to multiple inclusions.

\subsection{Multiple inclusions}
Let $r_1, \dots r_k$ be the radii of $k$ inclusions in the domain. We have $k+1$ superconducting domains and $k$ normal domains, each with their own first integral constant. The analysis from appendix \ref{critcurrent} carries over and we expect the inclusion domain's first integral constant $E_{I_k}$ to be approximately 0 for each $k$. This holds at the center of each respective inclusion, which each give different critical currents. However, when one is no longer satisfied, the system will no longer be satisfied and the global $j_c$ is determined by the lowest local $j_c$, which appears at the longest inclusion:
\begin{equation} \label{critcurrentgeneral}
j_c \approx \frac{1}{2} e^{-2 \max \limits_{k} r_k}.
\end{equation}

\subsection{Linear stability analysis of the stationary state} 
\begin{figure}[htb]
\includegraphics[width=\columnwidth]{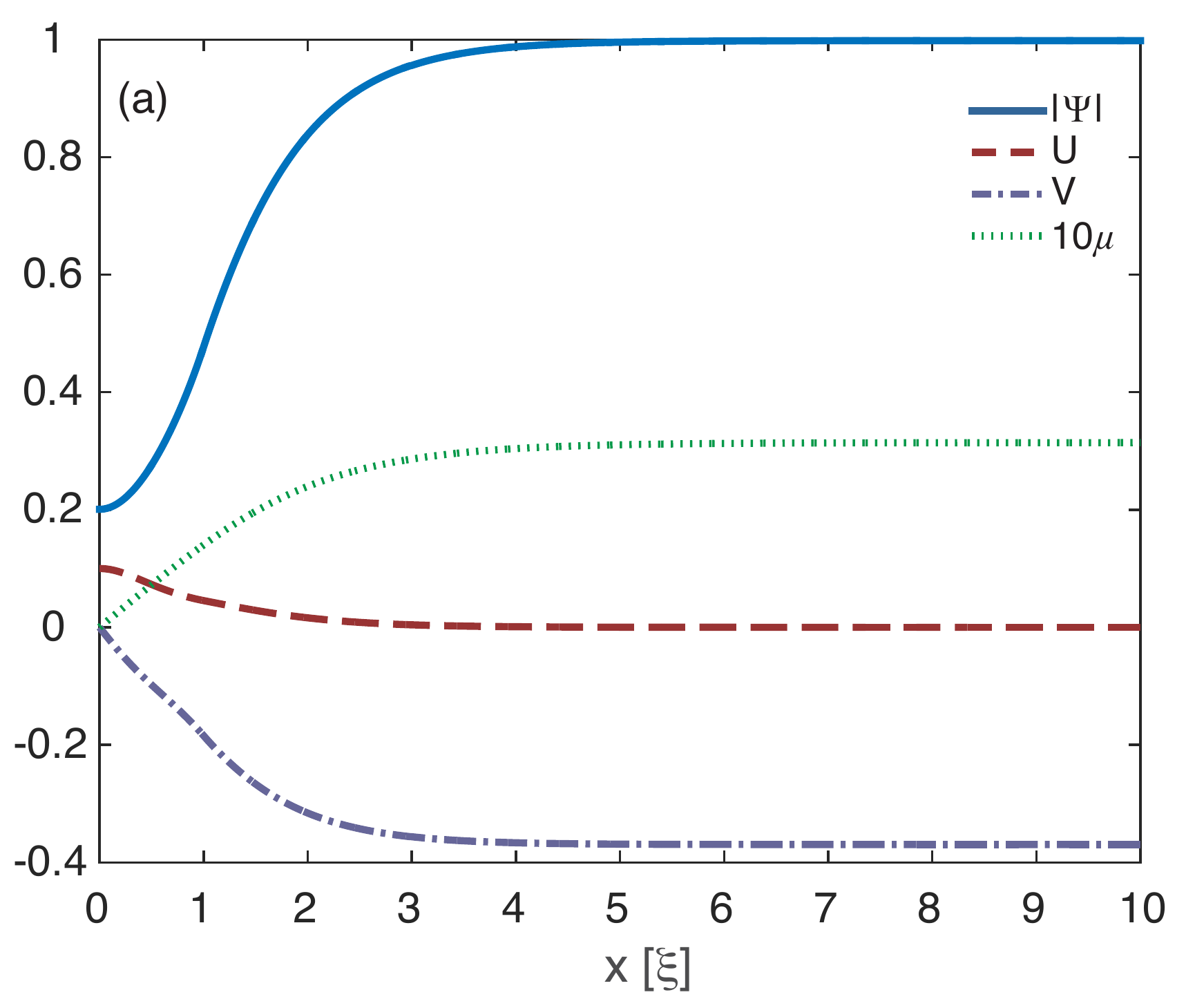}
\includegraphics[width=\columnwidth]{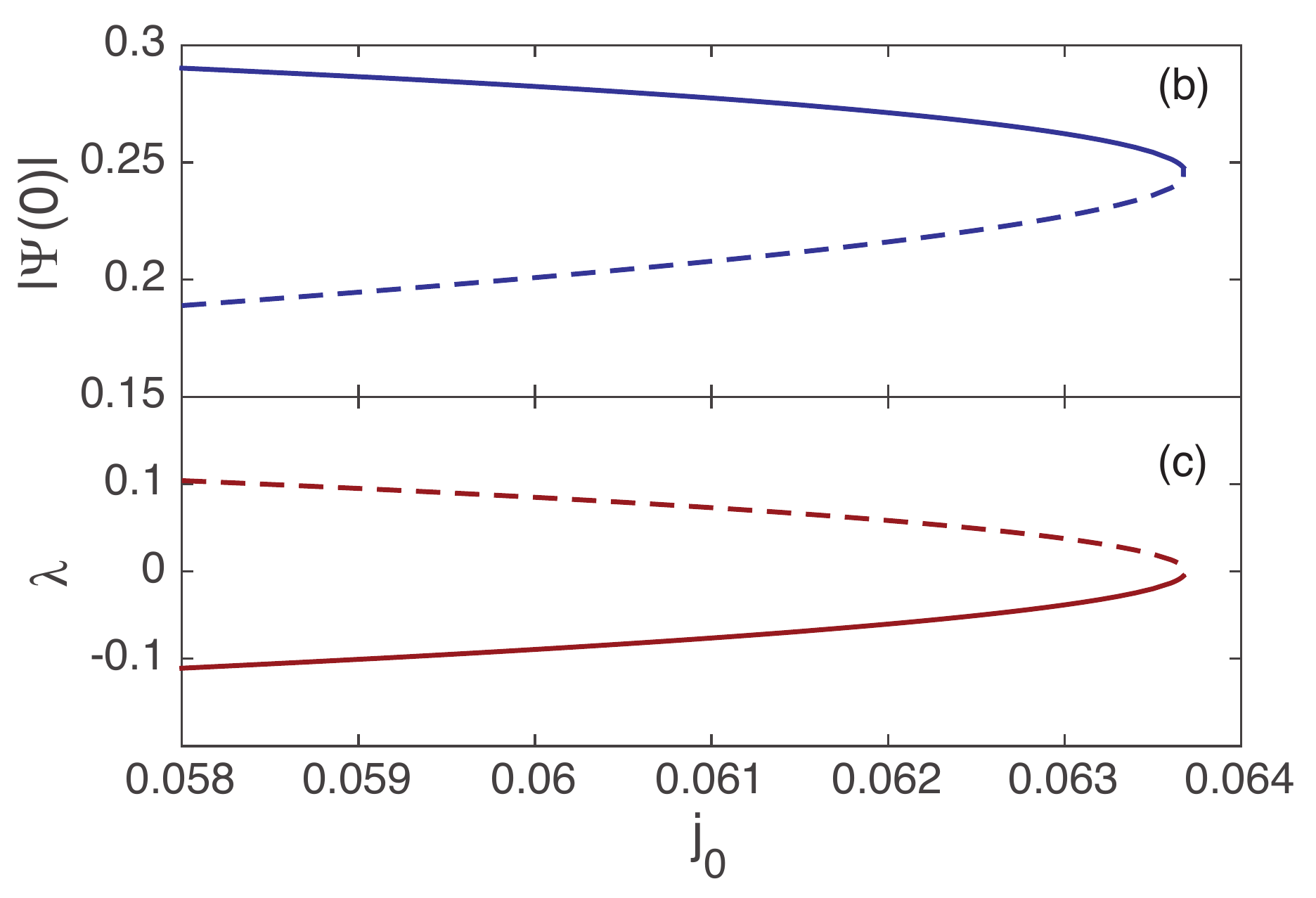}
\caption{(a) Amplitude $|\Psi|$ and linearized solutions $U,V, \mu$ with $j_0 = 0.061$, $r = 1$. plots (b) and (c) shows the value of $|\Psi(0)|$ and location of the smallest eigenvalue respectively, for stable (solid line) and unstable (dashed line) solutions of eqs. (\ref{stat2}), (\ref{linsys1})-(\ref{linsys3}) for varying current. At the critical current the stable and unstable stationary (i.e. superconducting) solutions merge and annihilate.}
\label{fig: linearized amplitude} 
\end{figure}

Consider now a perturbation $\eta$ of the stable state in the form $\Psi = (F + \eta) e^{i \phi}$. Inserting this into \eqref{GL1}-\eqref{GL2} and linearizing in $\eta$, we obtain with \eqref{stationary} and \eqref{current}
\begin{align*}
u \partial_t\eta & = \partial_x^2\eta + \left(\nu - (\partial_x\phi)^2 - 2F^2\right) \eta + \\
& \phantom{blah} i(2 \partial_x\phi \partial_x\eta +\partial_x^2 \phi \eta - u F \mu) - F^2 \eta^* \\
0 & = \Im(F \partial_x\eta + 2i F\partial_x \phi \eta + \partial_xF \eta^*) - \partial_x\mu.
\end{align*}
Separating $\eta(x,t) = (U + iV) e^{\lambda t}$ we obtain the following system (here $\lambda$ is the growth rate) 
\begin{subequations} \label{eq:linsys}
\begin{align}
\begin{split}
0 & = \partial_x^2U + \left(\nu - (\partial_x\phi)^2 - 3F^2 - \lambda\right) U - \\
& \phantom{blah} (2 \partial_x\phi \partial_xV + V \partial_x^2\phi) \label{linsys1}
\end{split} \\
\begin{split}
0 & = \partial_x^2V + \left(\nu - (\partial_x\phi)^2 - F^2 - \lambda\right) V + \\  \label{linsys2}
& \phantom{0 = } (2\partial_x\phi \partial_x U + U \partial_x^2\phi) - u F \mu \end{split} \\
\partial_x\mu & = F\partial_xV - V \partial_xF + 2 F U \partial_x\phi \label{linsys3}.
\end{align} 
\end{subequations}
This system along with \eqref{stationary}-\eqref{current} represents a 7 dimensional boundary-value eigenvalue problem which must be solved with appropriate boundary conditions. First, we note from \eqref{stationary} that replacing $x \to -x$ leaves the differential equation unchanged. This with the reflection symmetry implies that $F$ is an even function in $x$. This symmetry implies from \eqref{GL2} that $\partial_x \phi$ and $\partial_x \mu$ are even in $x$. Thus $x \to -x$ changes $\Psi \to \Psi^*$. The action of this must be retained in the linearization implying that $\eta(-x)$ and $\eta^*(x)$ are both solutions. Hence $U$ is even and $V$ is odd in $x$. Furthermore, by symmetry it suffices to solve the equations only on the half interval $(0,L/2)$ with the obtained natural boundary conditions from symmetry and the remaining conditions to be found by matching-shooting algorithm. To solve this we used a technique developed in \cite{tsimring1997localized,aranson1998surface}. In order to do so, we used a numerical matching-shooting solver for ODEs by beginning with a small domain (typically $L \sim 3$). We extracted the appropriate shooting boundary conditions and approximation for $\lambda$ and used these as guesses for a larger system size. Iterating this process, we continued to $L$ sufficiently large until the boundary conditions and $\lambda$ were not changing significantly. 
The results are plotted in Fig. \ref{fig: linearized amplitude}. We note here that $j_c \approx 0.0637$ obtained by the solver is only 6\% away from the value obtained through direct numerical solution of the Ginzburg-Landau model. The step size used in the dynamic simulations were much larger ($\Delta x = 0.05$ compared to shooting solver with $\Delta x = 0.001$) and each had an associated numerical error. Therefore, $j_c \approx 0.0637$ is more accurate. We checked if the error is independent of the solvers by analyzing the dynamic simulations $j_c$ as a function of $\Delta x$ in appendix \ref{numerical_jc}. We found that as $\Delta x \to 0$, we approached a similar value to that found from shooting. Thus, from Fig.~\ref{fig: linearized amplitude} one sees that at the critical current, when stable ($\lambda<0$) and unstable ($\lambda>0$) solutions merge and annihilate, the corresponding linear system becomes degenerate. At the critical point it possesses two zero eigenvalues $\lambda_{1,2}=0$. This degeneracy is taken into account through weakly nonlinear analysis.

\section{\label{sec:level4}Analysis of time-periodic solutions for  $j_0 > j_c$}
\begin{figure}[htb]
\includegraphics[width=\columnwidth]{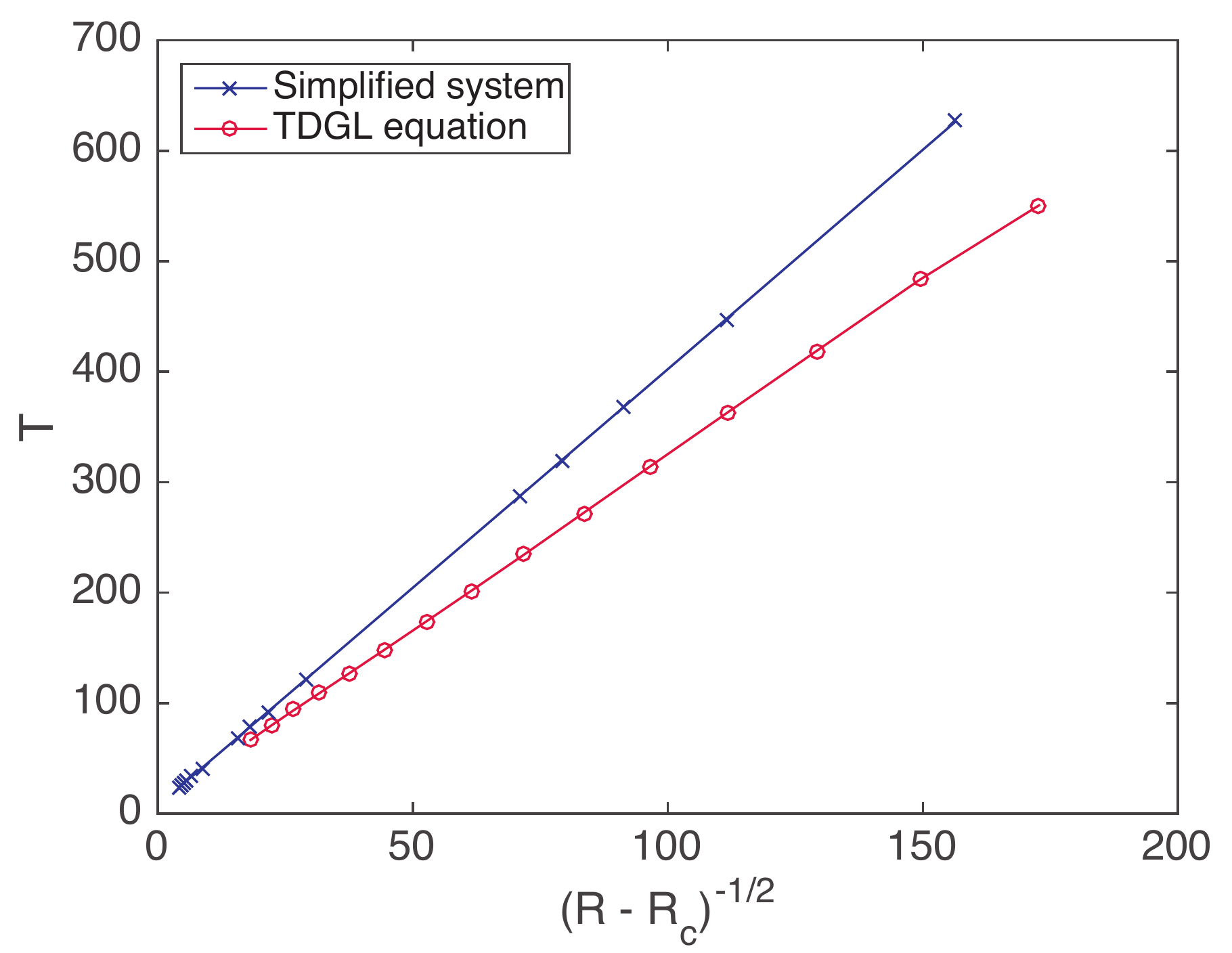}
\caption{IPB analysis with $L = 20$ and $r = 1$. The critical current $j_c \approx 0.067$ was obtained via stable state calculation from Section \ref{sec:level2}. The simplified system derived in section \ref{sec:level7} from weakly nonlinear analysis at $\gamma = -0.13$ with $c_\text{IP} \approx -0.565$ also exhibits an IPB. As expected, period  $T \propto \frac{1}{\sqrt{R - R_c}}$ near the bifurcation point in both cases. Here $R$ is current $j$ in the TDGLE and parameter $c$ in the simplified system.}
\label{fig:IPreal}
\end{figure}

When the current is above the critical threshold, the above analysis breaks down. Numerical simulations indicate that the superconductor exhibits oscillations in the order parameter, where phase slips are now present (i.e. $|\Psi(0,t)| = 0$ for some $t$). In figure \ref{fig:IPreal} we have estimated the period of oscillation $T$ as a function of $j_0 - j_c \ll 1$. Numerical simulations indicate that the period $T \sim O(|j_0 - j_c|^{-1/2})$, which is indicative of an infinite-period bifurcation (IPB) at the point $j_0 = j_c$. In general for a bifurcation parameter $R$ (e.g. current $j$) the period of oscillations $T \sim O(|R - R _c|^{-1/2})$ for $|(R/R_c) - 1| \ll 1$ for an IPB\cite{strogatz2014nonlinear}. We can see from figure \ref{fig:IPreal} that an IPB is occurring at the critical value. In section \ref{subsec:d2}, we show that for $u \gg 1$, we also observe hysteresis, behavior which is typical of a homoclinic bifurcation, a different mechanism through which a limit cycle can be destroyed\cite{strogatz2014nonlinear}.

\begin{figure}[htb]
\includegraphics[width=\columnwidth]{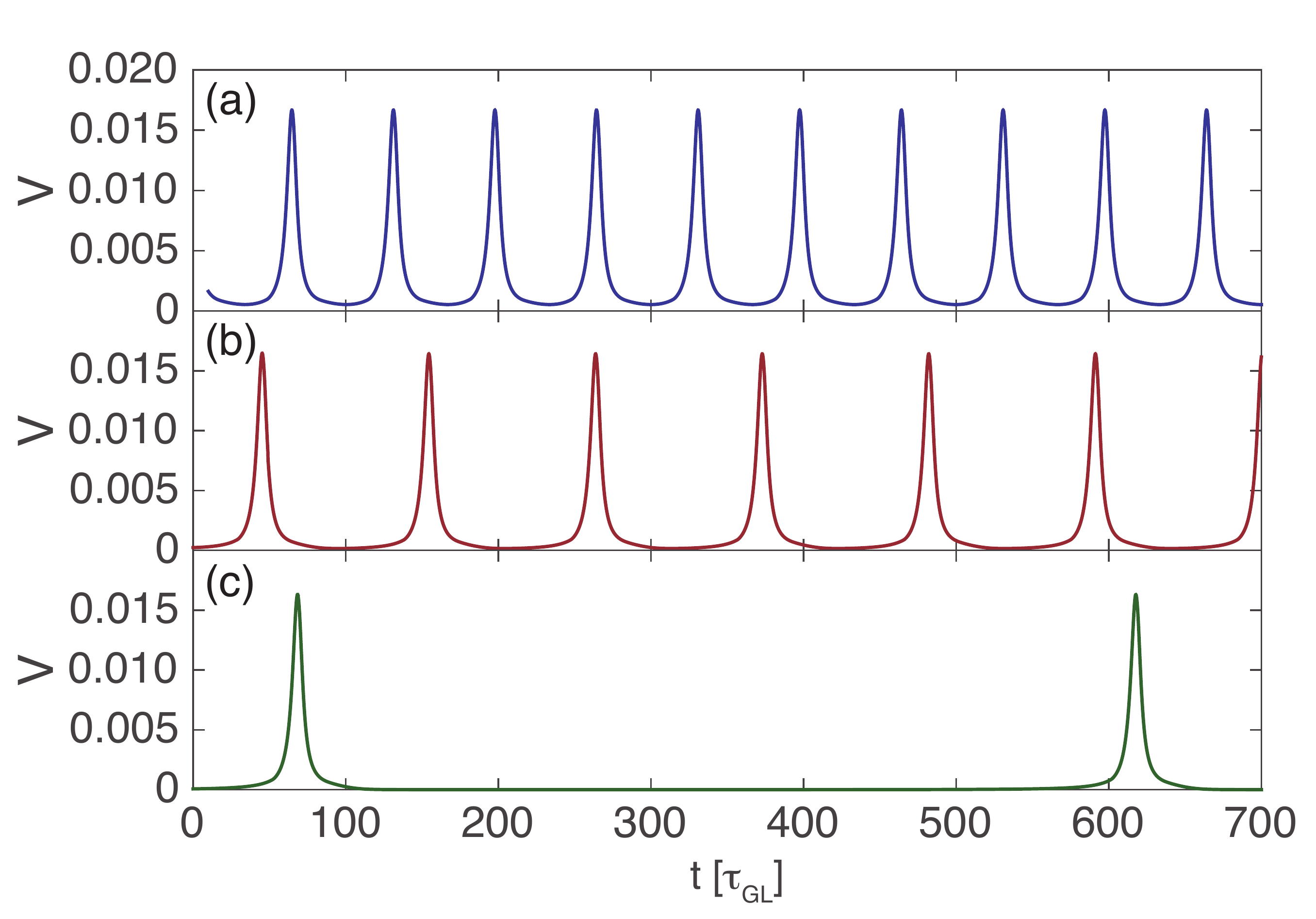}
\caption{Plots (a)-(c) show dependence of voltage vs time above the critical current where $j_0 = 1.045j_c$, $j_0 = 1.015j_c$ and $j_0 = \left(1+ 10^{-6}\right) j_c$, respectively and $j_c \approx 0.067$. System size $L = 20$, with an inclusion $r = 1$ in the center.}
\label{fig:VoltageTimePlot}
\end{figure}

\begin{figure}[htb]
\includegraphics[width=\columnwidth]{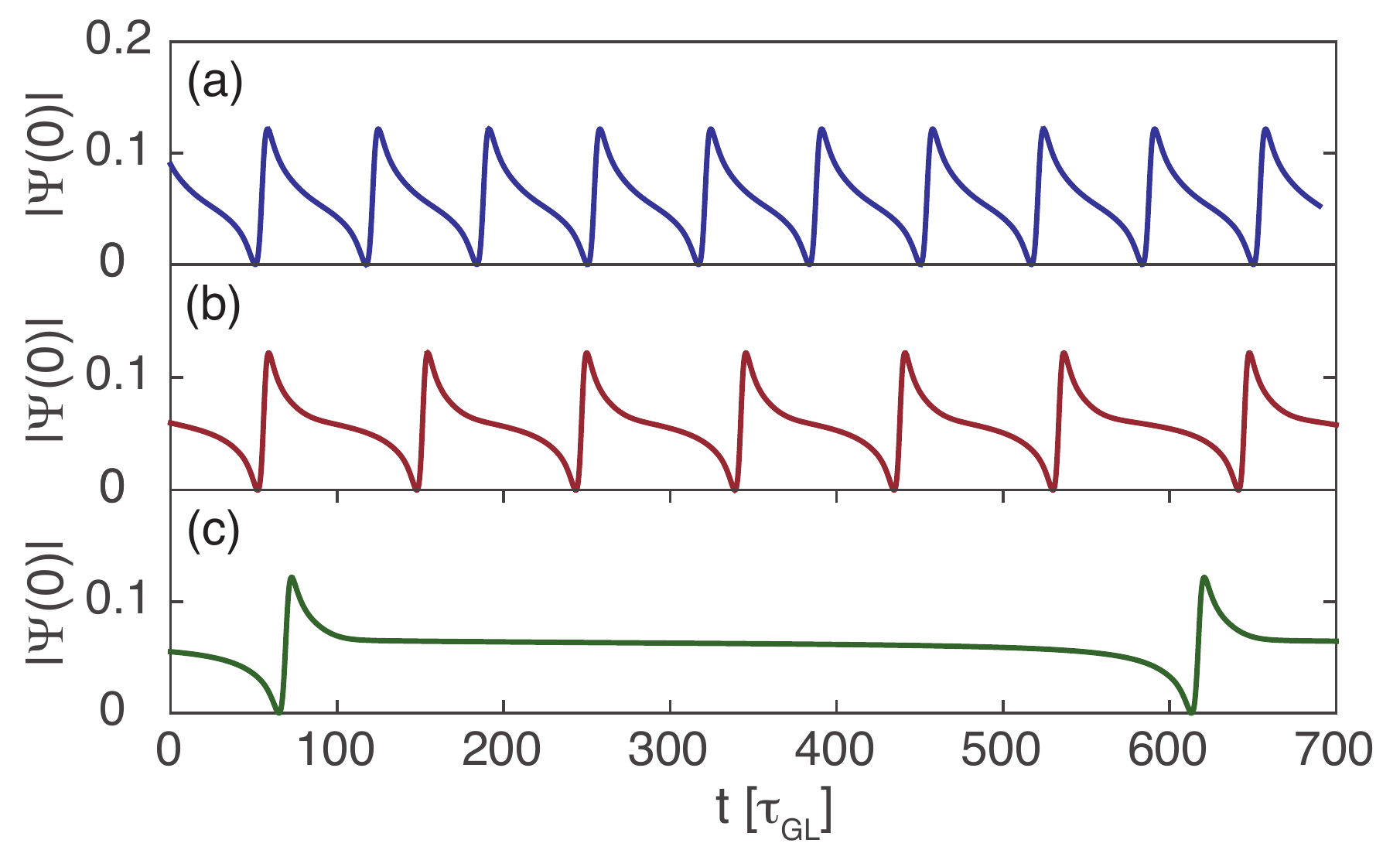}
\includegraphics[width=\columnwidth]{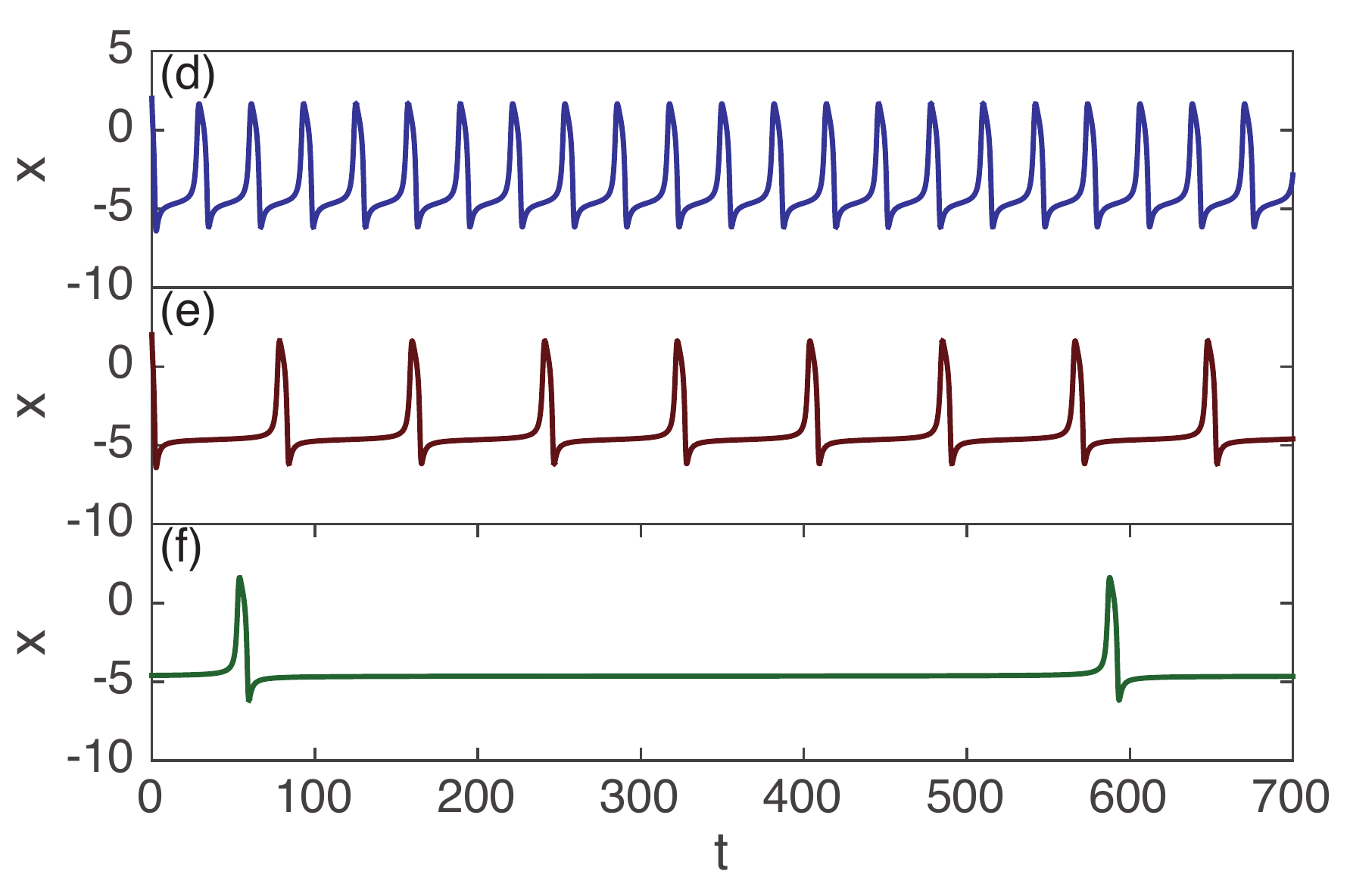}
\caption{Plots (a)-(c) show dependence of $|\Psi(0)|$ vs time above the critical current where $j_0 = 1.045j_c$, $j_0 = 1.015j_c$ and $j_0 = \left(1+ 10^{-6}\right) j_c$, respectively and $j_c \approx 0.067$. System size $L = 20$, with an inclusion $r = 1$ in the center. Plots (d)-(f) correspond to the simplified system Eqs. (\ref{fullODE}) where $\gamma = -0.13$ with $c_\text{IP} \approx -0.565$ is the IPB threshold, with $c = 0.955c_{\text{IP}}$, $c = 0.985c_{\text{IP}}$ and $c = \left(1 - 10^{-5} \right)c_{\text{IP}}$.} \label{fig:timevsxpsi}
\end{figure}

Figure \ref{fig:VoltageTimePlot} shows time-voltage curves  for $j_0 > j_c$. One clearly sees the period diverging as we approach the critical value. To calculate the current-period relationship, we ramped the current from an initial amount (typically $j_{\text{init}} < j_c$). If the system was stationary for a certain number of iterations, we increased the current. Once the system started oscillating, we calculated peaks in voltage, while skipping the first few to account for system equilibration. Then we averaged over the remaining peaks to obtain the period. We then used linear extrapolation to find the new current. For example, at the $n^{\text{th}}$ step, we have the current $j_n$ and corresponding period $T_n$. Let $m_n = \Delta T_n/\Delta j_n$, then suppose we want to find the current corresponding to a new period $T_{n+1} = (1+\alpha) T_n$, with $\alpha > 0$. This is given by $j_{n+1} = j_n + \frac{\alpha T_n}{m_n}$. Figure \ref{fig:timevsxpsi} shows a similar period divergence of the oscillations of $\Psi$ and the simplified model (see section \ref{sec:level7}).

\section{\label{sec:level5}Weakly nonlinear analysis}
We now extract a coupled ODE system, which exhibits two dynamical possibilities. In the case $j_0 < j_c$, we show that the stationary (fixed) solution is stable, while in the opposite case, a stable limit cycle exists. It is of course possible that a bistability region can exist, which would lead to hysteric effects. Such effects have been observed in homogeneous superconductors \cite{weber1991dissipative,PhysRevB.84.094527,PhysRevB.87.174516}. For large $u$, we have also observed hysteric I-V curves and we show that our extracted system contains both possibilities. The process is standard and is broken into these steps:
\begin{itemize}
\item Find stationary (basic) state $\Psi_0 = Fe^{i \phi}$ (it is already shown in Fig. \ref{fig: linearized amplitude}) 
\item Perturb solution and solve linearized system.
\item Extract weakly nonlinear effects from orthogonality condition.
\item Show that certain conditions allow for a stable limit cycle to exist.
\end{itemize}
Though standard, the difficulty in this problem is that the basic state and linearization cannot be solved in closed form. Though we can approximate it to a certain degree, its region of validity is dependent on the radius of the inclusion $r$, the current $j_0$ and to a smaller extent, the system size $L$. Indeed, it is impractical to obtain it numerically since the solutions are sensitive to these choices. However, our analysis will assume that these are all known {\emph a priori} and proceed through the framework. The simplified system is then obtained generally, and we show that the system exhibits the appropriate behavior for certain values in parameter space.

We expand Eqs. (\ref{NGL1})-(\ref{NGL3}) near the stationary solution and near the critical point $j_0 = j_c + \epsilon$ with $\epsilon \ll 1$. The first order solution will be given by $\Psi_0 = F e^{i \phi}$ (since $K = 0$, there is no electric potential in the super conducting state), in fact the initial transient would show exponential decay of $K \to e^{-\bavg{|\Psi|^2}t}$ and so $\mu = 0$ as expected. Let $\Psi = (F + \eta) e^{i \phi}$, where $\eta$, and time will now both slowly vary and be controlled by a small parameter $0 < \delta \ll 1$, whose size will be related to $\epsilon$. The proper scaling will be determined from the ODE for $K$. Based on numerical simulations, we assume $K = O(\delta^2)$. We claim that we may regard $K$ as constant in the relevant order of the perturbation method by the following argument. The perturbation $\eta$ at first order is highly localized inside the inclusion and from this we argue that
\begin{align*}
\avg{|\Psi|^2}  & = \frac{1}{L} \int_0^{L} F^2 + 2 F(\eta + \eta^*) + O(\eta^2) \, dx \\
& \approx \frac{1-j_0^2}{L} (L - r) + \frac{1}{L} \int_0^r F^2 + 2 F (\eta + \eta^*) \, dx \\
& \approx 1 - j_0^2 + O \left(\frac{r}{L}\right).
\end{align*}
For $ L \gg r$, we can regard $\avg{|\Psi|^2}$ as a constant. In a similar way all averaged quantities in the voltage equation can be neglected in the large superconductor domain limit. This analysis shows that the time-dependence of the voltage is slaved to the behavior of the order parameter $\Psi$. Therefore, we set $K$ to a constant by
\begin{equation} \label{Kconst}
K = \frac{\epsilon}{1- j_0^2} + O \left(\frac{r}{L}\right).
\end{equation}
From this, we extract the relation $\epsilon = \alpha \delta^2$ where $\alpha = \pm 1$. The linearized system at $\epsilon = 0$ has a degenerate eigenvalue as was shown previously in Fig. \ref{fig: linearized amplitude}. Therefore we expand $\eta(x,\tau) = A(\tau) \eta_1(x) + \sqrt \delta [B(\tau) \eta_2(x) + z_1(x,\tau)] + \delta z_2(x,\tau)$ where $\mathcal L \eta_1 = 0$, $\mathcal L \eta_2 = \eta_1$ and $\mathcal L$ is the linear operator from \eqref{eq:linsys}. Using orthogonality conditions, we arrive at the coupled system
\begin{align} \label{ODE}
\begin{split}
u A_\tau & = B + c_1 A^2 \\
u B_\tau & = c_2 AB + c_3 A^3,
\end{split}
\end{align}
where the coefficients $c_k$ can be found through evaluating the integrals (see appendix \ref{perturbation}). We will show in section \ref{sec:level6} why we chose to not include the constant $K$ at this order. The general behavior is only captured correctly at $\epsilon = 0$. When $\epsilon \neq 0$ (i.e. $K \neq 0$) we do not see a saddle-node bifurcation. To correct for this deficiency, higher order terms will be included. However, we can still gain some insight by analyzing this simplified system first.

\section{\label{sec:level6}Dynamical System Analysis}
We begin with \eqref{ODE} by making a dimensionless system to analyze it more easily. We introduce the dimensionless variables
\[ x = \frac{A}{L_A}, Y = \frac{B}{L_B}, t' = \frac{t}{u L_t}. \]
Inserting this into the system and defining the characteristic variables
\[ L_A = \frac{1}{c_2 L_t}, \quad L_B = \frac{1}{c_2 L_t^2}, \]
we arrive at the dimensionless system
\begin{align} \label{nondimODE}
\begin{split}
\dot X & = Y + a X^2 \\
\dot Y & = X Y + b X^3,
\end{split}
\end{align}
where $a\equiv c_1/c_2$ and $b \equiv c_3/c_2^2$. The characteristic scale for time is arbitrary and is a consequence of the degeneracy in the system. The culprits are the $X^2$ term and $XY$ terms whose combination of characteristic scales simultaneously vanish.

\subsection{Fixed points and stability} 
There is only one fixed point located at the origin, provided that $a \neq b$. In this case there is a family of non-isolated fixed points along the parabola $Y = - a X^2$, however this case is not physical so we omit it. Next, we note the symmetry $t \to -t$ and $X \to -X$ of \eqref{nondimODE}, which implies that the linearized center located at the origin is robust. We wish to see if this system exhibits closed orbits. The system is conservative if $a = -1/2$. In this case, a first integral can be obtained  
\[ H(X,Y) = \frac{1}{2} Y^2 - \frac{1}{2} X^2 Y - \frac{1}{4} b X^4. \]
This has closed orbits provided that $b < -1/2$. So now that we have established the existence of closed orbits, we seek to gain insight if $a \neq -1/2$. We replace $Y$ via the transformation
\[ Y = \frac{U}{2a+1} - a X^2, \]
and rescale $X \to \frac{X}{2 a + 1}$ and obtain
\begin{subequations} \label{oneparam}
\begin{align}
\dot X & = U  \label{oneparam1} \\
\dot U & = U X + \frac{b-a}{(2a+1)^2} X^3 \equiv U X + \gamma X^3.  \label{oneparam2}
\end{align}
\end{subequations}
This leaves us with one independent parameter $\gamma$. We have already analyzed the case where $a = -1/2$ which, if $b < -1/2$ corresponds to $\gamma \to -\infty$ and has a family of closed orbits. If $b > -1/2$ then $\gamma \to \infty$ and we know this does not have closed orbits. Therefore, there must be some critical value of $\gamma$ where this behavior changes. We seek a solution of \eqref{oneparam} of the form $X = \tilde C t^{-1}$ with $\tilde C$ to be determined. Plugging this into the equation gives the condition
\[ \tilde C = \frac{1 \pm \sqrt{1 + 8 \gamma}}{2 \gamma}. \]
These two solutions form a saddle-type connection only when they are equal which occurs at $\gamma_c = -1/8$ or in the original coefficients
\[ b_c = -\frac{1}{8}(2 a_c - 1)^2. \]

\begin{figure}[htb]
\includegraphics[width=\columnwidth]{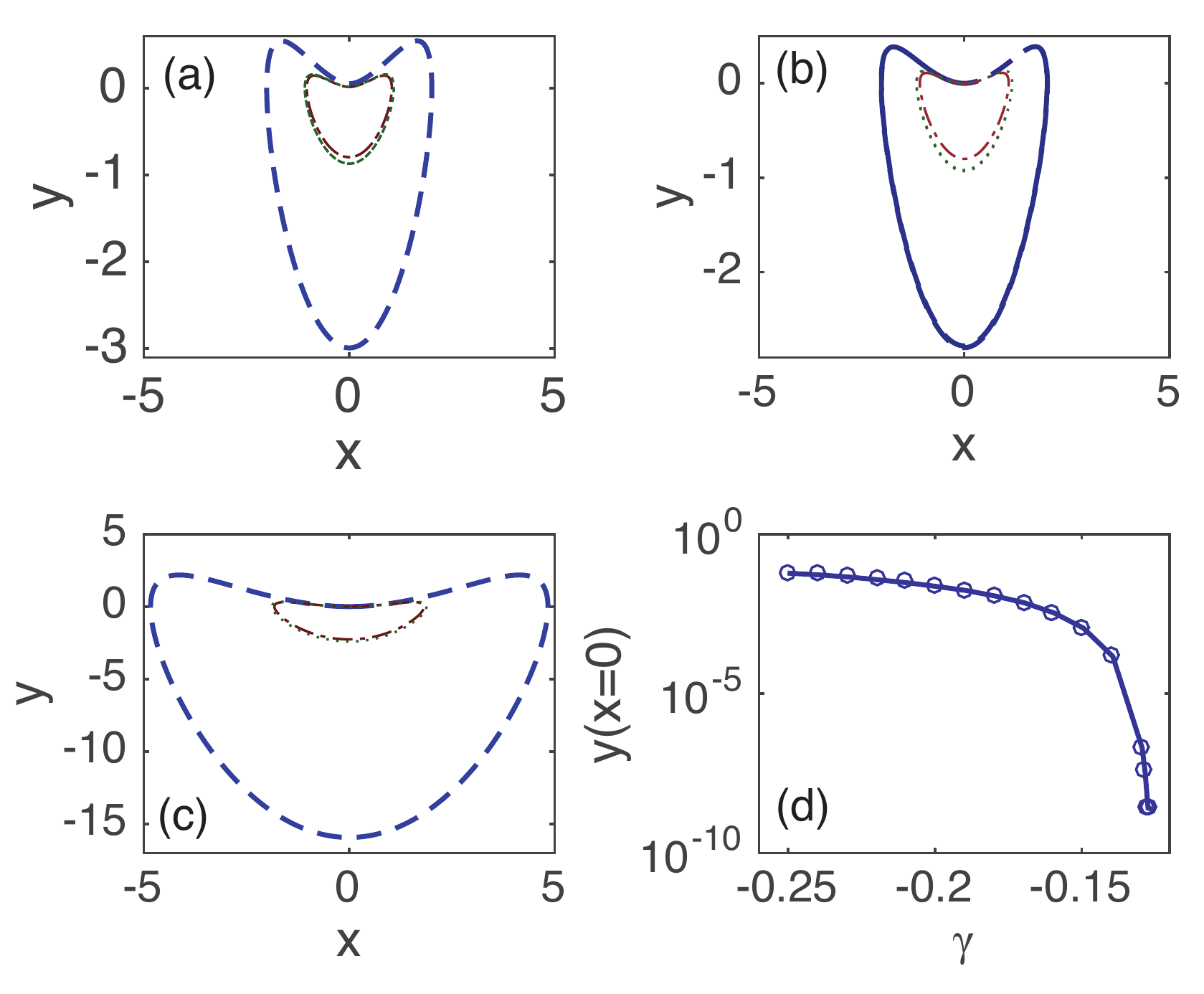}
\caption{Plots (a)-(c) represent the solutions to \eqref{oneparam} in the phase plane $(X,Y)$ with $\gamma = -0.25,-0.15,-0.13$, respectively. There is a dimple near the origin where the trajectories are being squeezed down due to the homoclinic orbit at $\gamma_c = -1/8$. In plot (d), we display this dimple as a function of $\gamma$ by taking 150 initial conditions and taking the average maximum.}
\label{fig: Gamma Curves}
\end{figure}

This critical curve separates closed orbit solutions in the $(a,b)$ parameter space. We have shown that the simplest (first order) system obtained, demonstrates a saddle-type infinite period bifurcation, however this creates an infinite family of closed orbits and a unique stable limit cycle is not obtained. The bottleneck is created near the origin (see Fig \ref{fig: Gamma Curves}). Additionally, it does not have a saddle-node bifurcation which we expect to occur at $j_0 = j_c$. We note also that introducing $K$ at this order, which adds a nonzero constant term to the second ODE would still only have one fixed point and a constant at this order would destroy the degeneracy (and also any closed orbits) in a degenerate Hopf-type bifurcation when that constant crosses through zero. This should be corrected by including the next higher order cubic terms which will saturate and force the system to select one unique closed orbit.

The bottleneck created near the emergence of the saddle-node bifurcation is apparent in both the physical and simplified system (see figure \ref{fig:timevsxpsi}). Note that the time scales need not be the same and careful treatment of the parameters in the simplified system (see section \ref{sec:level4}) would lead to the relation between the GL time and the time scale of the simplified system.

\section{\label{sec:level7}Full Dynamical System}
We modify the system to include the next order cubic terms. In principle, we could obtain the next order terms by continuing the perturbation expansion, however, we chose to include the generic next higher order terms $X^3, X^2 Y, X Y^2$, and so on. We then found that the removal of some cubic terms e.g. $X Y^2 , Y^3$ slightly shifts the transitions boundaries but does not qualitatively change the bifurcation sequence. Therefore, we chose to keep the following system for our analysis:
\begin{subequations} \label{fullODE}
\begin{align}
\dot X & = Y + a X^2 + w_1 X^3 \label{fullODE1}\\
\dot Y & = X Y + b X^3 + c + w_2 X^2 Y \label{fullODE2},
\end{align} 
\end{subequations}
where we have introduced the new coefficients $c, w_1, w_2$. We will enforce $w_1, w_2 < 0$ to ensure the phase flow cannot escape to infinity, which would be a nonphysical state for this system.

\subsection{Analysis}
The fixed points cannot be found analytically in general since the equation involves a quintic polynomial. Instead we look to find the two critical curves which correspond to our system. We wish to find a saddle-node bifurcation curve and an infinite-period bifurcation as the current is varied. The saddle-node bifurcation involves the merging and annihilation of the stable and unstable stationary solutions. An infinite-period bifurcation is a saddle-node bifurcation which occurs on the limit cycle in the phase plane \cite{strogatz2014nonlinear}.

We first find the fixed points of \eqref{fullODE}. Using \eqref{fullODE1}, we obtain  $Y^* = - (X^*)^2(a + w_1 X^*)$, which leads to the quintic equation
\[ f(X) \equiv w_1 w_2 X^5 + (w_1 + a w_2) X^4 + (a-b)X^3 - c = 0. \] 
A saddle-node bifurcation occurs provided that $f(X^*) = f'(X^*) = 0$. The curve exists only if $X^*$ is real which leads to the requirement that
\[ b \ge a -\frac{4(w_1 + a w_2)^2}{15 w_1 w_2}. \]
To motivate our choice of parameters, we write this in terms of $\gamma$
\[ \gamma \ge -\frac{4 w_1}{15 w_2} \left(\frac{\frac{w_2}{w_1}a + 1}{2a+1}\right)^2. \]
If we set $w_2 = 2 w_1$ we can eliminate $a$ from the dependence on $\gamma$. Thus, we have that the saddle-node bifurcation exists only if $\gamma \ge - \frac{2}{15}$.

Writing the quintic now with $a = -1$ allows us to cast the quintic function solely in terms of $w_1, \gamma \text{ and } c$.
\[ 2 w_1^2 X^5 - w_1 X^4 - \gamma X^3 - c = 0. \]
The saddle node bifurcation then occurs along the curve
\[ c_\text{SN}(X^*) = \frac{1}{5} (X^*)^3 \left[2 \gamma + w_1 (X^*)^2 \right], \]
where $X^*$ is given by
\[ X^* = \frac{1 \pm \sqrt{1 + \frac{15}{2} \gamma}}{5 w_1}. \]

The Jacobian of this system is
\[ J = \left[\begin{matrix} 2a X^* + 3 w_1 (X^*)^2 & 1 \\ Y^* + 3b (X^*)^2 + 2 w_2 X^* Y^* & X^* + w_2 (X^*)^2 \end{matrix} \right]. \]
A necessary condition for a Hopf bifurcation to occur is for a (un)stable spiral to change stability. This occurs when the trace of the Jacobian $\tau = X^*[2a+1 + (3w_1 + w_2) X^*] = 0$ and the determinant $\Delta > 0$. For our analysis  this implies that $X^* = 0$ or $X^* = (5 w_1)^{-1}$. Of course our fixed point $X^*$ must also satisfy the quintic equation. Inserting this gives a necessary condition and curve in $(\gamma,c)$ space for a Hopf bifurcation
\[ c_\text{Hopf} = -\frac{1}{125 w_1^3} \left(\gamma + \frac{3}{25} \right), \text{ or } c_\text{Hopf} = 0. \]
The determinant is
\[ \Delta = -\frac{1}{125 w_1^2}(2 + 15 \gamma). \]
Thus, the first Hopf bifurcation curve exists only when $\gamma < -2/15$. The second Hopf bifurcation is more complicated since $\Delta = 0$ and so nonlinear terms are important. The existence of that curve was found numerically.

\subsection{Phase Diagram} \label{subsec:phase diagram}
In general, this system has many different ways in which a limit cycle is destroyed. Numerical experiments indicate that this can occur via a Hopf, cycle bifurcation, infinite period or homoclinic bifurcation. Slightly changing the parameters can change which bifurcation we obtain. From the preceding section, we motivated the choices $w_1 = -0.05, w_2 = -0.1, a = -1$ to keep our parameter space $(\gamma, c)$. This leads to a generalized phase diagram of section \ref{sec:level6}.
\begin{figure}[htb]
\includegraphics[width=\columnwidth]{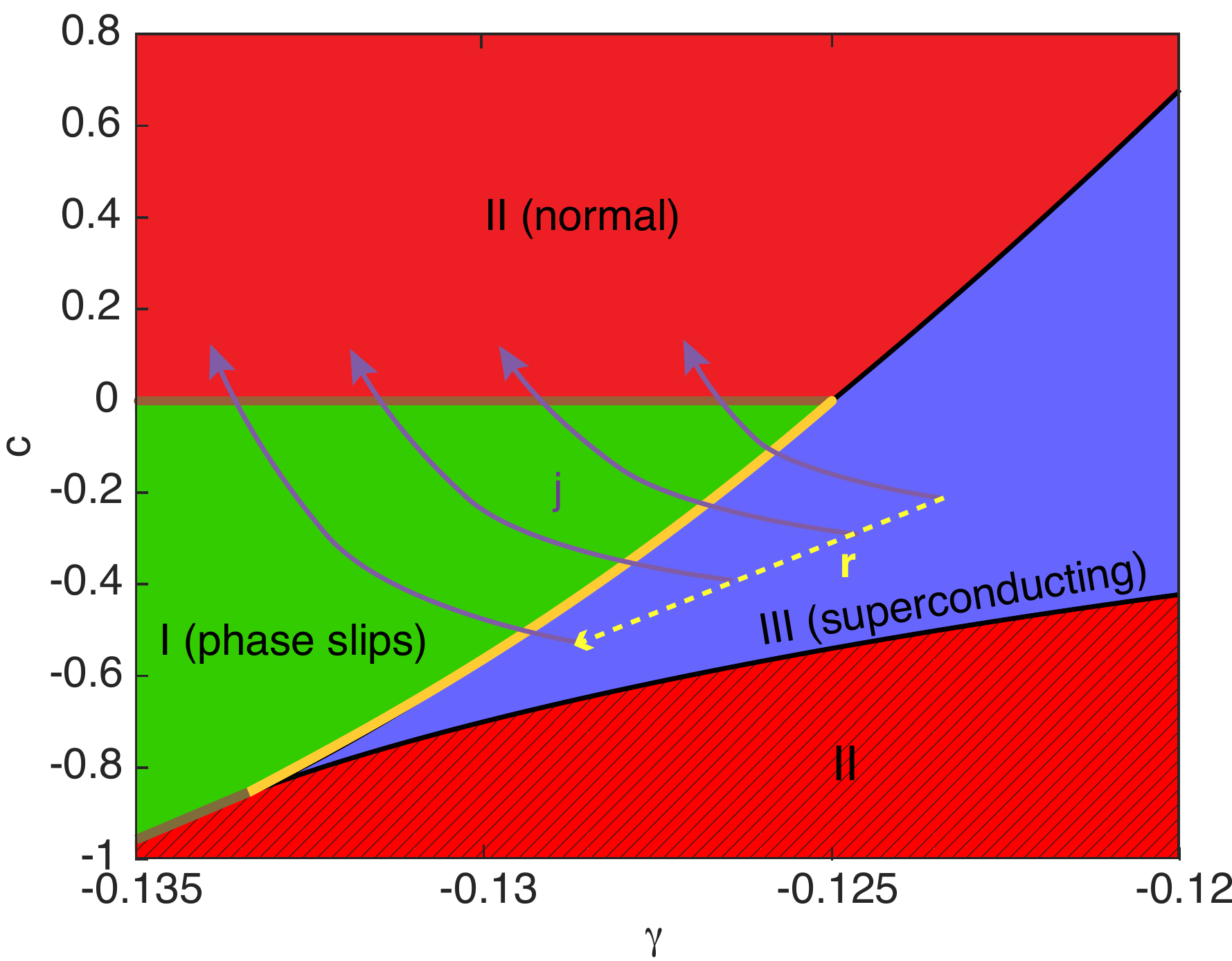}
\caption{Phase diagram with $a = -1$, $\gamma = b + 1$, $w_1 = -0.05$, $w_2 = -0.1$. There is a stable limit cycle, i.e. periodic phase slips,  (green) only in region I. Region II has one stable fixed point and region III has three fixed points. The saddle-node bifurcation (SNB) is boundary of the superconducting region. There is an IPB occurring along the yellow line. Possible trajectories in phase space are mapped with purple lines and the dashed yellow line corresponds to increasing $r$. Note that this phase diagram does not have a bistability region (with $u \gg 1$, we observed hysteresis, see section \ref{subsec:d2}).}
\label{fig:phaseplane1}
\end{figure}
The Hopf and saddle-node bifurcation curves of figure \ref{fig:phaseplane1} were obtained analytically. The IPB curve $c_{\text{IP}} = c_{\text{IP}}(\gamma_{\text{IP}})$ was found numerically and for comparison is compared to the observed physical limit cycle in figures \ref{fig:IPreal} and \ref{fig:timevsxpsi}. Additionally, it was found numerically that the HB in region III, did not exhibit the birth of a stable limit cycle. Possible trajectories of the superconductor through this phase diagram is shown with purple lines.

\begin{figure}[htb]
\includegraphics[width=\columnwidth]{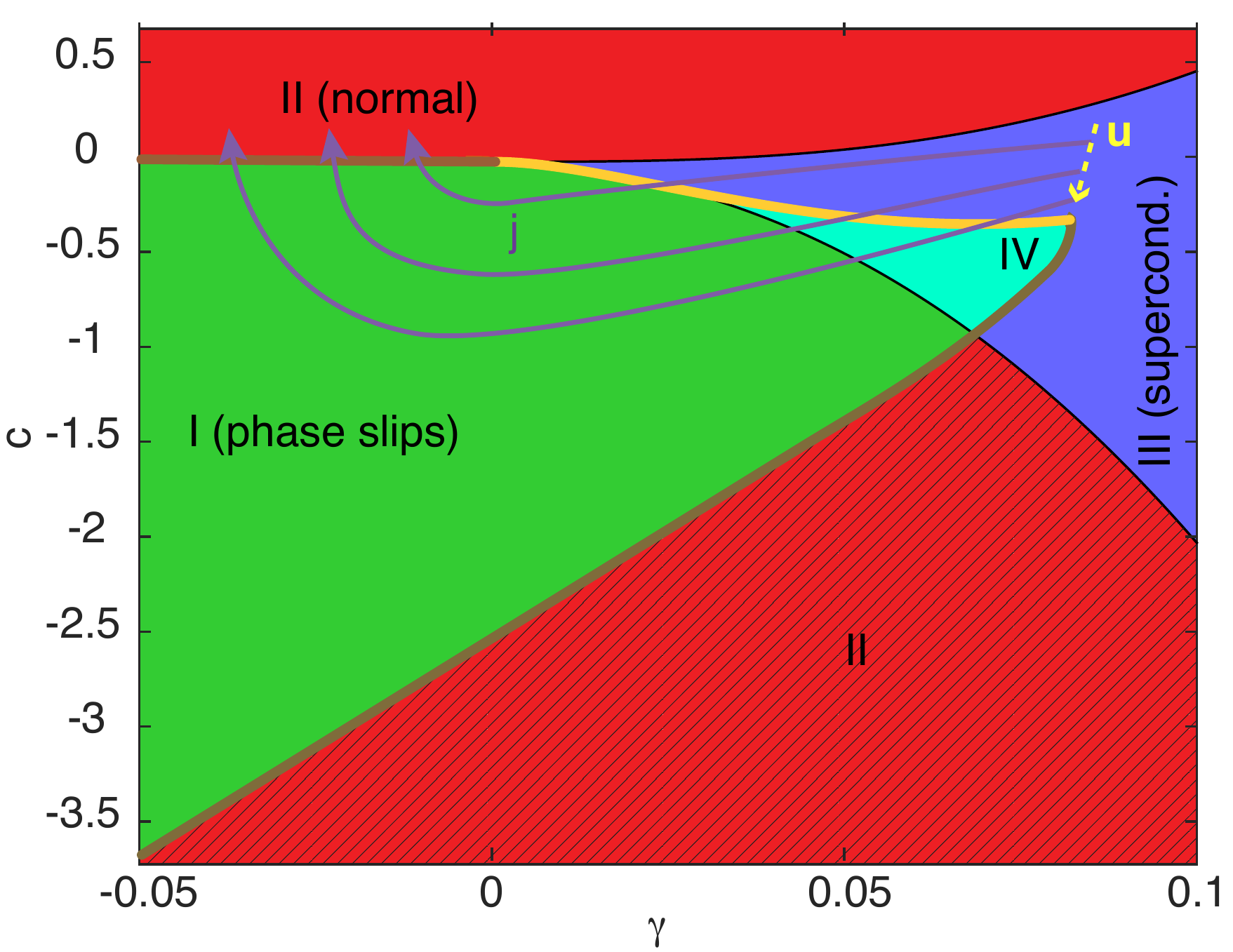}
\caption{Phase diagram with $a = -1$, $\gamma = b + 1$, $w_1 = -0.09$, $w_2 = -0.08$. There is a stable limit cycle (green) in region I. Region II has one stable fixed point and region III has three fixed points. Region IV is a bistability region where a limit cycle and distant attractor coexist. The limit cycle is destroyed along the yellow line via a homoclinic bifurcation (a saddle point moving towards the limit cycle), and the dashed yellow line corresponds to increasing $u$. This homoclinic bifurcation line eventually merges with the SNB line (boundary of region III) and becomes an IPB (similar to Fig. \ref{fig:phaseplane1}).}
\label{fig:phaseplane2}
\end{figure}

A more generic phase diagram with $w_2 \neq 2 w_1$ is given in figure \ref{fig:phaseplane2}. Here, both an IPB and homoclinic bifurcation can destroy the limit cycle. The existence of the homoclinic bifurcation changes the morphology of the phase diagram to now include a bistability region in which the limit cycle (phase slips) and fixed point (superconducting state) coexist. This is particularly encouraging since we also found hysteresis for $u \gg 1$ (see section \ref{subsec:d2}). Possible trajectories of the superconductor through this phase diagram is shown with purple lines.

\section{Discussion} \label{sec:discussion}

\subsection{Sensitivity to temperature} \label{subsec:d1}
To test the sensitivity of these phase slips to small thermal noise we modified \eqref{NGL} to include a small random noise term uniformly distributed between $[-T_f, T_f]$ at each point in space. Numerical simulations indicate that the system is stable to small fluctuations. The qualitative change is the existence of finite voltage in the superconducting state, however the critical current at which phase slips begin is unchanged.

\subsection{Effect of parameter $u$} \label{subsec:d2}
\begin{figure}[htb]
\includegraphics[width=\columnwidth]{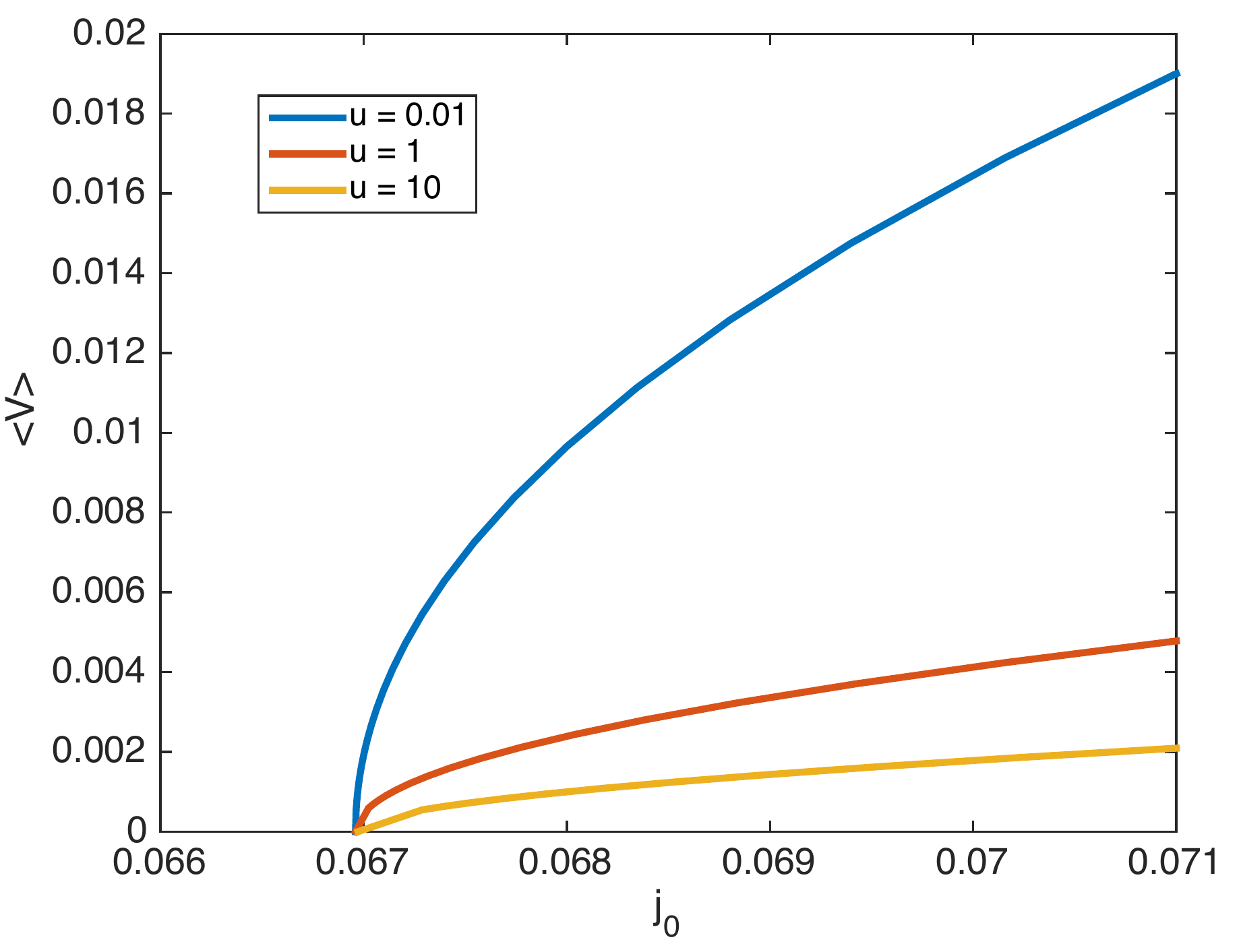}
\caption{I-V curve with different $u = 0.01, 1 ,10$, (color online). The critical current does not change, however the slope as $j_0 \to j_c$ increases as $u \to 0$. Additionally, $j_c = j_r$ (the reentrance current) for all $u$ shown (no hysteresis).}
\label{fig:u_eff}
\end{figure}

\begin{figure}[htb]
\includegraphics[width=\columnwidth]{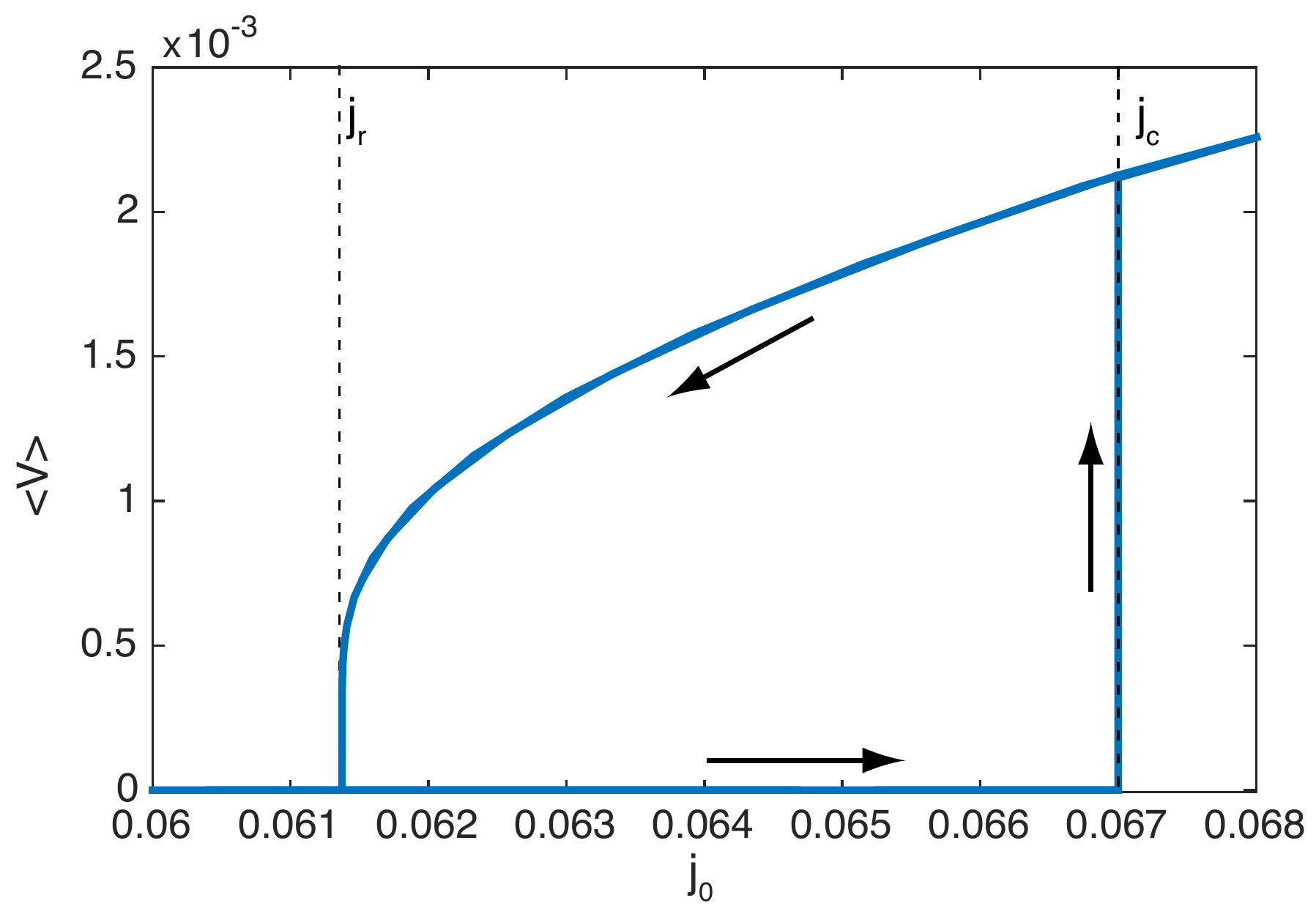}
\caption{I-V curve for $u = 100$. Hysteresis is present, the saddle-node bifurcation still occurs at $j_c \approx 0.067$, however $j_r \approx 0.0614$ below which the superconducting state reappears.}
\label{fig:u_100}
\end{figure}

The parameter $u$ characterizes the penetration of the electric field. In homogeneous wires, it has been found that hysteresis of the phase slip state exists for finite domains with $u \gg 1$ \cite{PhysRevB.87.174516}. We analyzed $u = 0.01, 1,10,100$ with $L = 20$ and $r = 1$ (see figure \ref{fig:u_eff}). Another important quantity not yet discussed is that of the retrapping current $j_r$. The authors of [\onlinecite{PhysRevB.87.174516}] discuss the effect of $u$, numerically simulating the GL equation and finding a curve separating the hysteresis region of the I-V curve through some length dependent critical curve $u_{c2}(L)$. For our simulations of weak links, $u$ small (for $r = 1$, $u < 30$ is small enough), $j_r = j_c$. However for $u \gg 1$, $j_r < j_c$, this leads to hysteresis in the I-V curve (see figure \ref{fig:u_100}).

\subsection{Physical quantities in simplified system} \label{subsec:d2}
The phase diagram is in $(\gamma, c)$ space. We can relate the important physical quantities $u,r, j_0,L$ to $\gamma, c$ by using appendix \ref{perturbation}. The coefficient $c$ is strongly affected by the parameter $u$ and the current $j_0$. Consider $j_0 < j_c$ and $u \to 0$, then we know that there is no voltage (i.e. $K = 0$), and $\alpha = -1$. This implies that $c \sim -u \zeta^2$ for some $\zeta(r,j_0, L)$ for small $u$. At a significantly large enough $u$ we expect our initial trajectory to begin from a region in figure \ref{fig:phaseplane2} where hysteresis is possible. Increasing the current $j_0 > j_c$ switches $\alpha = 1$ and $K \neq 0$, as $j_0$ continues to increase, $F$ decreases and we expect $c$ to change sign as we continue to increase it, which explains our motivation for the direction of trajectories. Increasing $r$ lowers $j_c$ and so we expect the trajectories to spend more time in the phase slip state, which leads us to expect that $c$ decreases. A similar argument, leads us to assume the same holds for $\gamma$ (see figure \ref{fig:phaseplane1} and \ref{fig:phaseplane2}). 
The effects on $\gamma$ are more complicated for the current and probably non-monotonous in a general case. From physical arguments we know that the trajectories must begin in the superconducting state and move into the phase slip state via either an IPB or homoclinic bifurcation. Comparing this to the phase diagrams, we see that as $j_0$ increases, $\gamma$ must decrease. We also attempt to justify this from the terms in appendix \ref{perturbation}. We consider the scaling from section \ref{sec:level6}, which implied $b = c_3/c_2^2$. We noted that $F$ is decreasing as $j_0$ increases (where $F'$ is relatively unchanged). Again, employing appendix \ref{perturbation}, we see that $c_3$ is decreasing with the current since the positive terms involve $F$ and the negative terms involve $F'$. Finally we use the fact that $b = c_3/c_2^2$ to deduce that $b$ must be decreasing and since $\gamma = (b-a)/(2a+1)^2$, we see that $\gamma$ is also decreasing with the current.

\section{Conclusion} \label{sec:conclusion}
We have considered a weak-link superconductor qualitatively similar to other weak-link systems, but fundamentally different in mechanism. We demonstrated the existence of a superconducting state and a PSC periodic state separated by a critical current $j_c$. This current was calculated asymptotically and agrees very well with numerical simulations. We then extracted a coupled ODE system from the TDGL equations using weakly nonlinear theory and showed under certain choices of parameters, an infinite period bifurcation and homoclinic bifurcations can occur. This demonstrates that the dynamics of phase-slip behavior in weak links described by the TDGL equations can be correctly captured by a simpler system of two coupled ordinary differential equations.

Further research is to extend this analysis to two dimensions. We anticipate additional transitions from phase slips occurring instantly inside the weak link to a more complicated dynamic regime involving phase slips and nucleation of vortex pairs, similar to that in [\onlinecite{weber1991dissipative}]. Another interesting generalization is to include disorder in the transverse direction inside the weak link. Possibly, some of the vortices will be pinned in the weak link. It may. in turn, lead to further suppression of the critical current. 

The work was supported by the Scientific Discovery through Advanced Computing (SciDAC) program funded by U.S. Department of Energy for computations. The Office of Science, Advanced Scientific Computing Research and Basic Energy Science, Division of Materials Science and Engineering for analysis.

\section{Appendix}

\appendix

\subsection{No voltage in the superconducting state} \label{nomu}
We begin by multiplying \eqref{GL1} by $\Psi^*$ and we differentiate \eqref{GL2} with respect to $x$. This gives
\begin{align}
u(i |\Psi|^2 \mu + \Psi^* \partial_t \Psi) & = \Psi^* \partial_x^2 \Psi + [\nu(x) - |\Psi|^2] |\Psi|^2 \label{mu1} \\
0 & = \Im{(\Psi^* \partial_x^2 \Psi)} - \partial_x^2 \mu \label{mu2}.
\end{align}
Taking the imaginary part of \eqref{mu1} and substituting this result into \eqref{mu2}, we obtain
\begin{equation} \label{mu3} \partial_x^2 \mu - u |\Psi|^2 \mu = u \Im{(\Psi^* \partial_t \Psi)}. \end{equation}
Far from the inclusion, all the applied current is supercurrent and so if $L \gg r$, we expect $j_0 = \Im{(\Psi^* \partial_x^2 \Psi|)_{x = \pm L}}$, which implies that $\partial_x \mu(\pm L) = 0$. Multiplying \eqref{mu3} by $\mu$ and integrating over the domain gives
\[ \int_{-L}^{L} \left[ (\partial_x\mu)^2 + u |\Psi|^2 \mu^2 \right) \, dx = \mu \partial_x \mu \bigg |_{-L}^{L} + u \int_{-L}^{L} \Im{(\Psi^* \partial_t \Psi)} \, dx. \]
Noting the boundary conditions for $\mu$ and the fact that $\partial_t \Psi = 0$ (stationary state), we see that $\mu \equiv 0$.

\subsection{Critical current calculation} \label{critcurrent}
 
We separate \eqref{stat2} by region (superconducting vs. normal metal) and then take the first integral to obtain the equations
\begin{align}
E_S & = (\partial_x F)^2 + F^2 + j_0^2 F^{-2} - \frac{1}{2} F^4, \quad && x \not \in I \label{FirstInt} \\
E_I & = (\partial_x F)^2 - C F^2 + j_0^2 F^{-2} - \frac{1}{2} F^4, && x \in I \label{SecondInt}.
\end{align}

Now, far from the inclusion (near the boundary of the superconductor), $F \to F_\infty$ a constant. Assuming the relevant approximation that $j_0 \ll 1$, we see that $F_\infty^2 \approx 1 - j_0^2$. Inserting this into \eqref{FirstInt}, implies that $E_S \approx \frac{1}{2} + j_0^2$. We now use the large $C$ approximation that $C \gg j_0^2 F^{-4}$. Proceeding, we obtain
\[ F_I(x) = K_1 e^{(|x|-r) \sqrt{C}}, \]
where we have introduced the radius $r$ of the inclusion. Solving the outer region at first order is given by
\[ F_S(x) = \tanh \left(\frac{|x|- K_2}{\sqrt{2}} \right). \]
The two constants are determined by the continuity conditions at the boundary of the inclusion. By symmetry, we may analyze just one side of the boundary, then our conditions are
\begin{subequations}
\begin{align}
K_1 & = \tanh \left(\frac{r-K_2}{\sqrt 2} \right) \\
K_1 & = \frac{1}{\sqrt{2C}} \sech^2 \left(\frac{r - K_2}{\sqrt 2} \right).
\end{align}
\end{subequations}
Solving for $K_1$ and $K_2$, we obtain
\begin{subequations}
\begin{align}
K_1 & = \frac{1}{\sqrt{2C}} + O \left(\frac{1}{C} \right) \\
K_2 & = r - \frac{1}{\sqrt C} + O \left(\frac{1}{C} \right).
\end{align}
\end{subequations}
Note the identity $E_S - E_I = (1+C)F^2(r) \ge 0$. This implies that 
\[ E_I \approx j_0^2 - \frac{1}{2C} \ll 1. \]
Motivated by this, we assume that $E_I$ is a small parameter. At first order then $E_I = 0$ and looking at $x  = 0$ we see that
\[ E_I = 0 = -CF^2(0) + j_0^2 F^{-2}(0) - \frac{1}{4} F^4(0), \]
where the derivative has vanished by symmetry. Since $F$ is small in the inclusion, the last term can be neglected and we are left with $j_0 \approx \sqrt{C} F^2(0)$. This leads to Eq. \eqref{critcurrentapprox}.

\subsection{Numerical analysis of $j_c$} \label{numerical_jc}
\begin{figure}[htb]
 \includegraphics[width=\columnwidth]{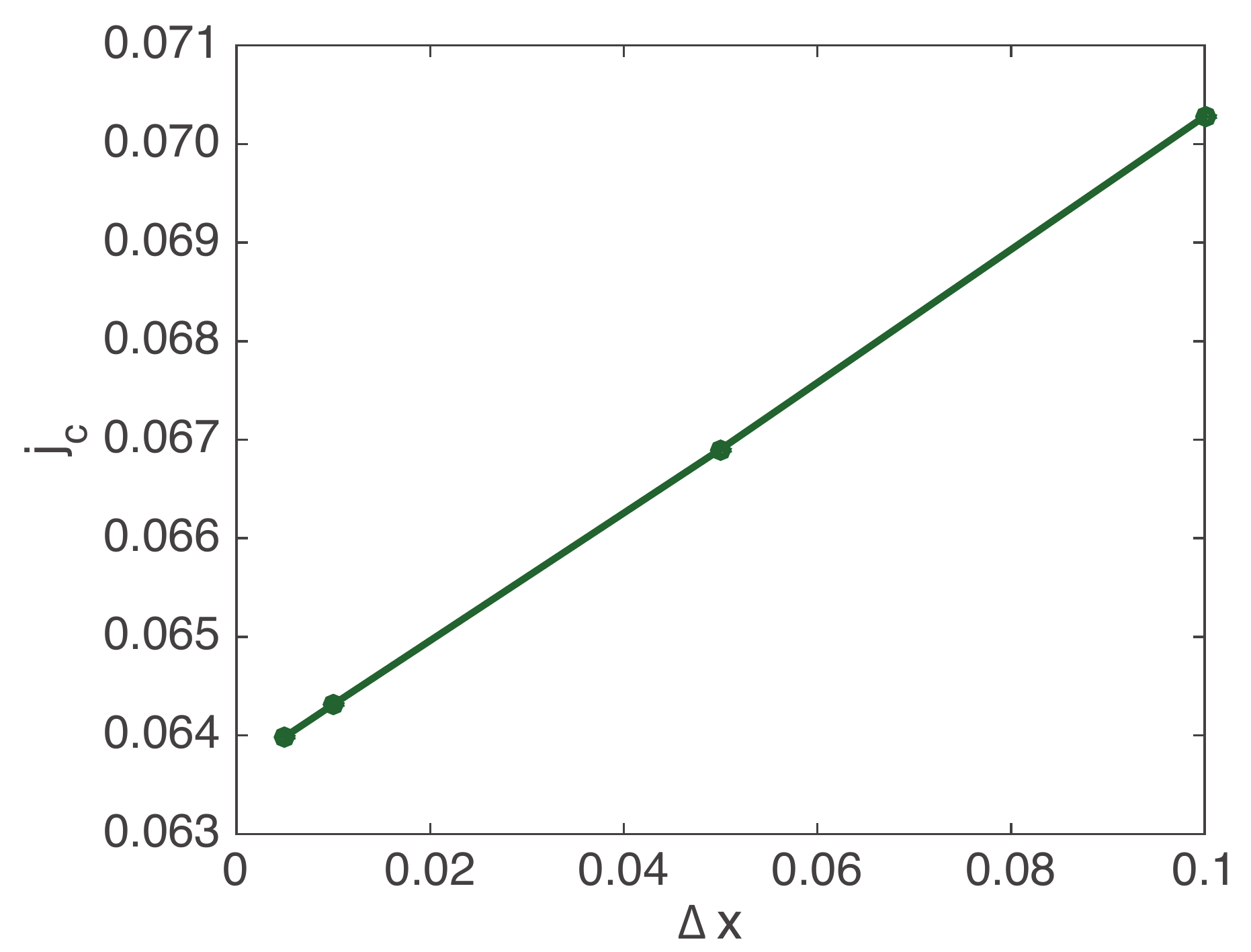}
 \caption{Convergence of $j_c$ as a function of $\Delta x$. As $\Delta x \to 0$, $j_c$ approaches the true value. Dynamic simulations took place with $\Delta x = 0.05$.}
 \label{numerical_jcFIG}
 \end{figure}

 To analyze the error associated with calculating $j_c$ numerically, we took $L = 20$ and varied $\Delta x$. The results are shown in figure \ref{numerical_jcFIG}. Assuming the error is linear, we extrapolate the critical current to be $j_c \approx 0.06366$, which is in excellent agreement with the linear system solved using the shooting method with $\Delta x = 0.001$.
 For fixed $\Delta x = 0.05$, we measured the sensitivity of $L$ on $j_c$ and found no significant change for $L \gg r$ (typically $L > 5r$ was sufficient).

\subsection{Weakly nonlinear calculation} \label{perturbation}
To obtain the weakly nonlinear system, we analyze near $j_0 = j_c + \epsilon$ where $|\epsilon| \ll 1$. Linearizing about the base state near $\epsilon = 0$ with $\Psi = (F + \eta) e^{i \phi}$. From before, we saw that $\epsilon = 0$ leads to a degenerate zero eigenvalue implying that the linearized system has a generalized eigenvector solution where $\mathcal L \eta_1 = 0$ and $\mathcal L \eta_2 = \eta_1$. We use Ansatz $\eta = A \delta \eta_1 + \delta^2 B (\eta_2 + z) + \delta^3 \zeta$ where $\eta_k = \left (\begin{matrix} U_k \\ V_k \end{matrix} \right)$ and $\epsilon = \alpha \delta^2$. Inserting this into \eqref{NGL1}--\eqref{NGL3} with the aid of mathematica and obtain at first order the ODE for $A$
\begin{widetext} 
\begin{eqnarray*}
\mathcal L z  & = & u \eta_1 \partial_\tau A - \left( \begin{matrix} B U_1 - A^2 \left[F(3U_1^2 + V_1^2) + u V_1 \int_{-L/2}^x (F' V_1 - F V_1' - 2 F \phi' U_1) \, ds \right] \\
B V_1 - A^2\left[ u U_1 \int_{-L/2}^x F V_1' - F' V_1+ 2 F \phi' U_1 \, ds  + F \left(2 U_1 V_1 + u \int_{-L/2}^x \{ \phi'(U_1^2 + V_1^2) + U_1 V_1' - U_1' V_1 \} \, ds \right) \right] \end{matrix} \right).
\end{eqnarray*}
At next order, we obtain the ODE for $B$ (where we have already projected onto the eigenvector)
\begin{eqnarray*}
u \partial_\tau B \langle U_1^\dag, U_2 \rangle & = & \bigg \langle U_1^\dag, A^3\left \{ u V_1 \int_{-L/2}^x [\phi' (U_1^2 + V_1^2) +  U_1 V_1' - U_1' V_1] \, ds - U_1^3 - U_1 V_1^2 \right \} -2 K F \phi' + \\ & \phantom{abcd} & A B \left [u V_2 \int_{-L/2}^x (2 F \phi' U_1 + F V_1' - F' V_1) \, ds - u V_1 \int_{-L/2}^x (2R \phi' U_2 + F V_2' - R'V_2) \, ds - 2R(3 U_1 U_2 + V_1 V_2) \right ] \bigg \rangle \\
u \partial_\tau B \langle V_1^\dag, V_2 \rangle & = & \bigg \langle V_1^\dag, - A^3 \left \{ U_1^2 V_1 + V_1^3 + u U_1\int_{-L/2}^x [\phi'(U_1^2 + V_1^2) + U_1 V_1' - U_1' V_1] \, ds \right \} - \\
& \phantom{abcd} & A B \bigg \{ U_2 \left[2 F V_1 + u \int_{-L/2}^x (F V_2' - F' V_2 + 2 F \phi' U_2) \, ds \right] + U_1 \bigg[2R V_2 + \\
& \phantom{abcd} & u \int_{-L/2}^x (F V_2'  - F' V_2 + 2 F \phi' U_2) \, ds \bigg] + u F \int_{-L/2}^x [U_2 V_1' -V_2 U_1' + U_1 V_2' - V_1 U_2' + 2\phi' (U_1 U_2 + V_1 V_2)] \, ds \bigg \} \\
& \phantom{abcd} & + K \left(2 F' - u F \int_{-L/2}^x F^2 \, ds \right) + u \alpha x F \bigg \rangle
\end{eqnarray*}
 \end{widetext} 
  
\bibliographystyle{apsrev4-1}

\end{document}